\documentclass{aa}  
\usepackage{graphicx}
\usepackage{txfonts}
\usepackage{aalongtable}
\begin{document}
\input{psfig}
%
% Next 5 lines define \simless and \simgreat: "less than or approximately
% equal to" and "greater than or approximately equal to".
\newbox\grsign \setbox\grsign=\hbox{$>$} \newdimen\grdimen \grdimen=\ht\grsign
\newbox\simlessbox \newbox\simgreatbox
\setbox\simgreatbox=\hbox{\raise.5ex\hbox{$>$}\llap
     {\lower.5ex\hbox{$\sim$}}}\ht1=\grdimen\dp1=0pt
\setbox\simlessbox=\hbox{\raise.5ex\hbox{$<$}\llap
     {\lower.5ex\hbox{$\sim$}}}\ht2=\grdimen\dp2=0pt
\def\simgreat{\mathrel{\copy\simgreatbox}}
\def\simless{\mathrel{\copy\simlessbox}}
% Next lines define "approximately proportional to"
\newbox\simppropto
\setbox\simppropto=\hbox{\raise.5ex\hbox{$\sim$}\llap
     {\lower.5ex\hbox{$\propto$}}}\ht2=\grdimen\dp2=0pt
\def\simpropto{\mathrel{\copy\simppropto}}

\title{VLT-FLAMES Analysis of 8 giants in
 the Bulge Metal-poor Globular Cluster NGC 6522: Oldest Cluster in the Galaxy?
\thanks{Observations collected both at the European Southern Observatory,
Paranal, Chile (ESO) programmes 71.B-0617A, 73.B0074A).
} }

\subtitle{Analysis of 8 giants in NGC 6522}

\author{
B. Barbuy\inst{1}
\and
M. Zoccali\inst{2}
\and
S. Ortolani\inst{3}
\and
V. Hill\inst{4}
\and
D. Minniti\inst{1,5}
\and
E. Bica\inst{6}
\and
A. Renzini\inst{7}
\and
A. G\'omez\inst{4}
}
\offprints{B. Barbuy}

\institute{
Universidade de S\~ao Paulo, IAG, Rua do Mat\~ao 1226,
Cidade Universit\'aria, S\~ao Paulo 05508-900, Brazil;
e-mail: barbuy@astro.iag.usp.br
\and
Universidad Catolica de Chile, Departmento de Astronomia y Astrofisica,
Casilla 306, Santiago 22, Chile;
e-mail: mzoccali@astro.puc.cl, dante@astro.puc.cl
\and
Universit\`a di Padova, Dipartimento di Astronomia, Vicolo
 dell'Osservatorio 2, I-35122 Padova, Italy;
 e-mail: sergio.ortolani@unipd.it
\and
Observatoire de Paris-Meudon,  92195 Meudon Cedex, France;
e-mail: Vanessa.Hill@obspm.fr, anita.gomez@obspm.fr
\and
Specola Vaticana, Vatican Observatory, V00120 Vatican City State, Italy
\and
Universidade Federal do Rio Grande do Sul, Departamento de Astronomia,
CP 15051, Porto Alegre 91501-970, Brazil; e-mail: bica@if.ufrgs.br
\and
Osservatorio Astronomico di Padova, Vicolo
 dell'Osservatorio 5, I-35122 Padova, Italy;
 e-mail:  alvio.renzini@oapd.inaf.it
}
 
   \date{Received; accepted }

% \abstract{}{}{}{}{} 
% 5 {} token are mandatory
 
  \abstract
   {NGC 6522 has been the first metal-poor globular cluster identified in the bulge
by W. Baade. Despite its importance, very few high resolution abundance analyses
of stars in this cluster are available in the literature.
The bulge metal-poor clusters may be  important tracers of
the early chemical enrichment of the Galaxy.}
   { The  main purpose  of this  study is the
determination  of metallicity and elemental  ratios in individual stars of NGC 6522. }
   {High resolution  spectra of 8 giants of  the bulge globular cluster NGC 6522
 were obtained  at the 8m VLT UT2-Kueyen  telescope with the  FLAMES+GIRAFFE
spectrograph.   Multiband $V,I,J,K_{\rm  s}$ photometry  was used to  
derive  effective temperatures as reference values. Spectroscopic parameters
are derived from \ion{Fe}{I} and \ion{Fe}{II} lines, and adopted
for the derivation of abundance ratios.}
   { The  present analysis   provides a
metallicity  [Fe/H] = $-1.0\pm0.2$.   The $\alpha$-elements  Oxygen, Magnesium and
Silicon show  [O/Fe]=+0.4$\pm$0.3, [Mg/Fe]=[Si/Fe]= +0.25$\pm$0.15,  whereas  Calcium 
and  Titanium show shallower ratios of [Ca/Fe]=[Ti/Fe]=+0.15$\pm$0.15.
 The neutron-capture r-process element Europium appears to be
 overabundant by [Eu/Fe]=+0.4$\pm$0.4.
The neutron-capture s-elements La and Ba are enhanced by [La/Fe]=+0.35$\pm$0.2 and
[Ba/Fe]=+0.5$\pm$0.5. The large internal errors, indicating the large star-to-star variation
in the Ba and Eu abundances, are also discussed. }
   {The moderate metallicity combined to a blue Horizontal Branch (BHB),
are characteristics similar to those of HP~1 and NGC 6558, pointing to a population
of very old globular clusters in the Galactic bulge. Also, the abundance ratios in NGC 6522
 resemble those in HP 1 and NGC 6558. The ultimate conclusion is that the bulge is old,
and went through an early prompt chemical enrichment. }
\titlerunning{Abundances in 8 giants of NGC 6522}
\authorrunning{B. Barbuy et al.}
   \maketitle
%
%________________________________________________________________

\section{Introduction}

Metal-poor bulge field stars and clusters represent a crucial piece in the
puzzle of the Milky Way formation. 
NGC 6522 and its surrounding fields, located in the Large Sagittarius
Cloud, were observed by Baade (1946), that identified part of the Cloud
as a window reaching the nuclear bulge, since then called Baade's Window.
Baade concluded, for the first time, that its stellar population
is of type II.
Blanco \& Blanco (1984) and Walker and Mack (1986)  presented B,V data on NGC 6522,
and the latter authors concluded that NGC 6522 is moderately metal-poor.

Lee (1992) has shown that the metallicity distribution of RR Lyrae
variables in the Galactic bulge is more metal-rich than in the halo,
with a peak metallicity at [Fe/H]$\approx$-1.0. This 
is interpreted as an age effect, given that more metal-rich stars
are expected to populate the red horizontal branch (RHB), and only lower
mass (and older) stars would be bluer falling in the RR Lyrae gap.
This implies as well that the oldest stellar
population of the Galaxy is found in the Galactic bulge. 

Lee et al. (2007) showed all clusters with extended blue horizontal branches (EHB)
 are the most massive and  brightest globular clusters of the Milky Way, all of them
brighter than magnitudes M$_{\rm V}$ $<$ -7.   NGC 6522 is classified as having
a moderately extended EHB and estimated integral magnitude of  M$_{\rm V}$ = -7.67
(Harris, 1996,   updated    in
www.physics.mcmaster.ca/Globular.html), or -7.99 (Armandroff 1989),
and is therefore at the edge of the distribution of massive clusters.
Lee et al. (2007) suggested that NGC 6522 is among relics of the first building blocks
that first assembled to form the Galactic nucleus, and that are now observed
as relatively metal-poor EHB globular clusters.

Therefore, NGC 6522, together with 
 other bulge clusters such as HP 1 (Barbuy et al.
2006a) and NGC 6558 (Barbuy et al. 2007), could
be relics of primeval star-forming sub-systems that first formed
the Galactic center population.
This could have been achieved  both by
dissipational and dissipationless mergers, as has been predicted by
recent $\Lambda$CDM simulations of high-$\sigma$ peaks
(e.g. Diemand et al. 2005; Moore et al. 2006).

 The globular cluster  NGC 6522, also designated GCl 82,
C 1800-300 and Cl VDBH 256,  is located at 
J2000 $\alpha=18^{\rm h}03^{\rm m}34.08^{\rm s}$, $\delta=-30^{\rm o}02'02.3''$,
 and projected  at  4$^{\circ}$ from the
Galactic   center (l=1.0246$^{\circ}$,   b=$-3.9256^{\circ}$).  
It     is at a distance 
d$_{\odot}=6$ kpc  away from the  Sun, and  at ${\rm R_{GC}}= 2$ kpc from
the  Galactic  center  (Barbuy  et al.  1998).   

Terndrup et al. (1998) derived proper motions of 
$<$$\mu_l$$>$=1.4$\pm$0.2 mas yr$^{-1}$,
$<$$\mu_b$$>$=-6.2$\pm$0.2 mas yr$^{-1}$, and a mean radial
velocity of v$_{r}$ = -28.5$\pm$6.5 km s$^{-1}$, and concluded
that the cluster stays in the bulge.

Literature basic parameters  of NGC 6522 are gathered in
Table~\ref{tab1}. Minniti et al. (1995) presented a first K vs. J-K 
Colour-Magnitude Diagram (CMD) of NGC 6522.
 The  cluster has a  concentration parameter c=2.50, 
a core radius log  r$_{\rm c}$(') = 0.49,  
and a half-light radius of  log r$_{\rm h}$(')= 1.78 (Trager  et al. 1995).

Among the metal-poor clusters of the inner bulge, Terzan~4 ([Fe/H]=-1.6) has been
studied with high resolution infrared spectroscopy (Origlia \& Rich 2004),
revealing significant enhancement of the $\alpha$-elements. HP 1
([Fe/H]=-1.0) and NGC 6558 ([Fe/H]=-1.0) were studied with high resolution
spectroscopy in the optical (Barbuy et al. 2006a, 2007), showing shallow 
$\alpha$-element enhancements.
     
In this work  we present a detailed abundance analysis  of 8 stars  in
NGC 6522, based on high resolution spectra obtained
 with FLAMES+GIRAFFE at the ESO  Very 
Large Telescope  VLT-UT2 Kueyen, at Paranal. The detailed analysis is
 carried out using updated MARCS model atmospheres (Gustafsson et al. 2008).

   The observations are described  in Sect. 2.
  Photometric stellar parameters effective
temperature and  gravity  are   derived in  Sect.   3.    Atomic and
molecular data are reviewed in Sect.  4. Spectroscopic parameters are
derived in  Section  5 and   abundance ratios are   computed in  Sect.
6. A discussion is presented in Sect. 7 and conclusions are drawn in Sect. 8.

\begin{table}
\caption[1]{
 Parameters of NGC 6522 given in the literature. References: 1 Zinn 1985; 2 Armandroff 1989;
3 Harris 1996; 4 Rutledge et al. 1997a;  5 Rutledge et al. 1997b (given in Zinn \& West 1984 and Carretta
\& Gratton metallicity scales); 6 Terndrup \& Walker 1994;
7 Terndrup et al. 1998 
(A$_{\rm V}$=1.42 is given);
8 Piotto et al. 2002; 9 Kraft \& Ivans 2003; 10 Bica et al. 2006.  }
\begin{flushleft}
\begin{tabular}{llllll@{}llll}
\noalign{\smallskip}
\hline
\noalign{\smallskip}
\hline
\noalign{\smallskip}
E(B-V) &
[Fe/H] &
v$_r$ &
$(m-M)$ &
 d$_{\odot}$ &
ref.  \\
 &
 &
(km s$^{-1}$) &
 &
 (kpc) & \\
\noalign{\vskip 0.2cm}
\noalign{\hrule\vskip 0.2cm}
\noalign{\vskip 0.2cm}
0.45    & -1.44            &       +8   & 15.37  & 6.2    & 1 \\
---     & -1.44            & -3    & ---    & ---    & 2  \\
0.48    & -1.44            & -21.1      & 16.52  & 7.8    & 3 \\
0.50    & -1.44            & -18.7      & ---    & ---    & 4 \\
---    & -1.50/-1.21      & ---        & ---   & ---    & 5  \\
0.52    & -1.60             & ---        & ---    & ---   & 6 \\
0.39    & -1.28            & -28.5      & ---    & 7.3   & 7 \\
0.48    & -1.44            & ---        & ---    & 7.4   & 8 \\
---     &-1.44/-1.42/-1.35 & ---        & ---    & ---    & 9 \\
0.48    & -1.44            & ---        & ---    & 7.8    & 10 \\
\noalign{\smallskip} \hline \end{tabular}
\end{flushleft}
\label{tab1}
\end{table}

%____________________________ OBSERVATIONS ___________________________
%--- 2 ---
\section {The Data}

\subsection{Imaging}

$V$ and $I$ data of NGC 6522 and surrounding fields were collected
from the Optical Gravitational Lensing Experiment (OGLE) survey 
(Udalski et al. 1992, 1993).     
Target stars were cross identified in the 2MASS Point Source Catalogue
(Skrutskie et al. 2006 \footnote{
$\mathtt{http://ipac.caltech.edu/2mass/releases/allsky/}$})
and their $J$, $H$, and $K_{\rm s}$ magnitudes were used,
except for the fainter sample stars B-108, B-122, B-130, F-121,
for which no satisfactory identification in 2MASS data was found, 
their magnitudes being probably at the photometric limits of the 2MASS survey.

\begin{figure}[ht]
\psfig{file=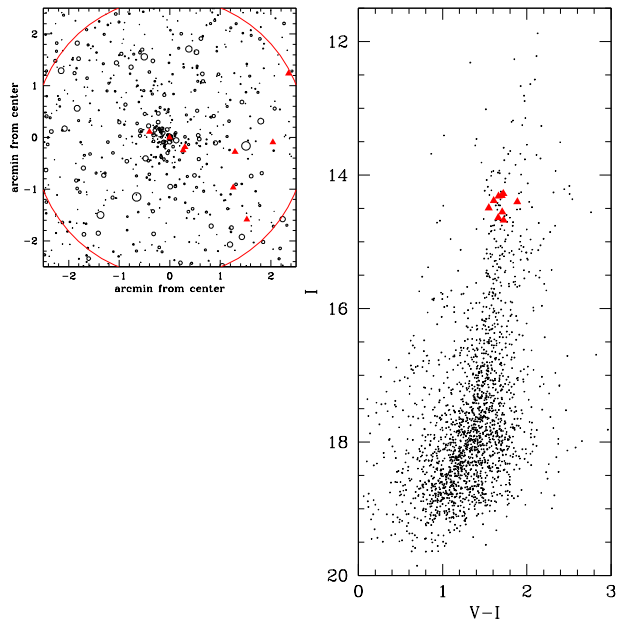,angle=0,width=9cm}
\caption {OGLE Colour Magnitude Diagram of Baade's Window, with
the sample stars of NGC 6522 overplotted.}
\label{cmd_ogle}
\end{figure}

The  location   of  target   stars  on   the OGLE  CMD   is  shown  in
Fig.~\ref{cmd_ogle}. Six of the target stars were identified in the field
of an ACS (Advanced Camera for Surveys) image retrieved from the archives
of the Hubble Space Telescope, as shown in Fig. \ref{manuds9}.

  The selected stars, their coordinates, magnitudes  and colours as 
 obtained from the OGLE catalogue,   together with 
the 2MASS designations (Skrutskie et al. 2006), coordinates
and VIJHK$_s$ magnitudes, are listed in Table~\ref{starmag}.
The star  IDs are adopted according  to
the fact that FLAMES+GIRAFFE has provided spectra for 134 stars in each of the
bright (B) and faint  (F)  stars selected in this  field.
The cluster members turned out to be star B8 (i.e., the 8$^{\rm th}$
spectrum of  the bright   sample), and  so forth, and    such
designations were kept.

\begin{table*}
\caption[1]{Identifications, positions, magnitudes, and 
dereddened  colours  adopting the reddening
law by Dean  et al. (1978)  and  Rieke \&  Lebofsky (1985). }
\begin{flushleft}
\tabcolsep 0.15cm
\begin{tabular}{ccccccccccccccccccc}
\noalign{\smallskip}
\hline
\noalign{\smallskip}
\hline
\noalign{\smallskip}
{\rm star} & 2MASS ID & $\alpha_{2000}$ & $\delta_{2000}$ & $V$ & $I$ & $J$ & $H$ & $K_{\rm s}$ &   $K_{TCS}$ 
& {\rm V-I$_{\rm J}$$_{\circ}$} & {\rm V-K$_{\rm TCS}$$_{\circ}$} & {\rm J-K$_{\rm TCS}$$_{\circ}$} & \cr
\noalign{\vskip 0.2cm}
\noalign{\hrule\vskip 0.2cm}
\noalign{\vskip 0.2cm}
B-008 &18034605-3000512 & 18:03:46.04 & -30:00:50.9 & 15.99 & 14.384 & 13.099 & 12.336 & 12.198 & 12.205 & 1.008 & 2.548 & 0.626 & \cr
B-107 &18033660-3002164 & 18:03:36.59 & -30:02:16.1 & 15.98 & 14.316 & 13.006 & 11.803 & 11.574 & 12.008 & 1.066 & 2.735 & 0.726 &  \cr
B-108 & ---             & 18:03:35.18 & -30:02:04.9 & 16.29 & 14.403 & ---    & ---    & ----   & ----   & 1.289 & ---   & ---   & \cr
B-118 &18034225-3003403 & 18:03:42.24 & -30:03:39.9 & 16.01 & 14.309 & 13.056 & 12.305 & 12.142 & 12.149 & 1.103 & 2.624 & 0.638 &\cr
B-122 & ---             & 18:03:33.34 & -30:01:58.3 & 16.00 & 14.280 & ---    & ---    & ---    & ---    & 1.122 & ---   & ---   & \cr
%B-122 &18033338-3001588 & 18:03:33.34 & -30:01:58.3 & 16.00 & 14.280 & 12.707 & 11.988 & 11.128 & 11.139 & 1.122 & 3.626 & 1.277 & \cr
B-128 &18034463-3002107 & 18:03:44.61 & -30:02:10.4 & 16.26 & 14.553 & 12.860 & 12.614 & 12.438 & 12.442 & 1.109 & 2.578 & 0.166 & \cr
B-130 & ---             & 18:03:41.00 & -30:03:03.0 & 16.30 & 14.640 & ---    & ---    & ---    & ---    & 1.062 &  ---  & ---   & \cr
%B-130 &18034102-3003036 & 18:03:41.00 & -30:03:03.0 & 16.30 & 14.640 & 12.850 & 12.307 & 10.634 & 10.649 & 1.062 & 4.418 & 1.889 & \cr
B-134 &18034117-3002218 & 18:03:41.16 & -30:02:21.6 & 16.04 & 14.493 & 13.286 & 12.768 & 12.563 & 12.569 & 0.949 & 2.233 & 0.455 & \cr
F-121 &18033640-3002221 & 18:03:36.41 & -30:02:19.8 & 16.40 & 14.676 & 12.586 & 12.065 & 11.729 & 11.584 & 1.126 & 3.580 & 1.135 &  \cr
\noalign{\smallskip} \hline \end{tabular}
\label{targets}
\end{flushleft}
\label{starmag}
\end{table*}

\begin{table*}
\caption[2]{Log of the spectroscopic observations carried out on 2003 May 7 (Julian Date 2452766),
2003 June 7, 11, 23, 26 (Julian Dates 2452797, 2452801, 2452813, 2452816) and 
2003 July 20 (Julian Date 2452840). Slit widths of 0.8'' and 1.0'' were used, depending on the
seeing value. The quoted seeing is the mean value along the exposures.}
\begin{flushleft}
\begin{tabular}{lllllllccccccc}
%@{}
\noalign{\smallskip}
\hline
\noalign{\smallskip}
\hline
\noalign{\smallskip}
Target  & setup & date & UT & exp   & Airmass  & Seeing & (S/N)/px  & ${\rm v_r^{obs}}$ & ${\rm Mean}$ &
${\rm v_r^{hel.}}$ & ${\rm Mean}$ \\
\noalign{\smallskip}
&  & &  &  (s) &   & ($''$) &   &${\rm km s^{-1}}$ &${\rm km s^{-1}}$ &${\rm km s^{-1}}$ &${\rm km s^{-1}}$ \\
\noalign{\smallskip}
\noalign{\smallskip}
\hline
\noalign{\vskip 0.2cm}
B-008 & HR13 & 07.05.03 & 05:37:22.3 & 5400  & 1.15 & 0.96$''$  & 128 & -30.088 & -29.5 & -9.03 & -19.8 \\
      & HR14 & 23.06.03 & 03:50:00.5 & 9900  & 1.025 & 0.68$''$ & 107 & -28.762 & & -28.82 &  \\
      & HR15 & 07.06.03 & 03:49:11.9 & 3600  & 1.115 & 0.57$''$ & 160 & -29.633 & & -21.52 &  \\
B-107 &HR13 & 07.05.03 & 05:37:22.3 & 5400  & 1.15  & 0.96$''  $ & 120 & -27.633 & -27.3 & -6.59 & -17.7 \\
 &HR14 & 23.06.03 & 03:50:00.5 & 9900  & 1.025 & 0.68$''$       & 120 & -26.527 & & -26.60 &  \\
 &HR15 & 07.06.03 & 03:49:11.9 & 5400  & 1.15 & 0.96$''$        & 194 & -27.751 & & -19.94 & \\
B-108  &HR13 & 07.05.03 & 05:37:22.3 & 5400  & 1.15  & 0.96$''$  & 110 & -33.111 & -32.7 &  -12.07 & -23.0 \\
 &HR14 & 23.06.03 & 03:50:00.5 & 9900  & 1.025  & 0.68$''$       & 106 & -33.605 & & -33.60 &  \\
 &HR15 & 07.06.03 & 03:49:11.9 & 5400  & 1.15   & 0.96$''$        & 100 & -31.244 & & -23.44 &  \\
B-118  &HR13 & 07.05.03 & 05:37:22.3 & 5400  & 1.15 & 0.96$''$  &  94 & -43.045 & -41.7 &  -22.00 & -32.0 \\
 &HR14 & 23.06.03 & 03:50:00.5 & 9900  & 1.025  & 0.68$''$       &  92 &  -42.293 & & -42.35 & \\
 &HR15 & 07.06.03 & 03:49:11.9 & 5400  & 1.15   & 0.96$''$        &  92 & -39.806 & & -31.99  & \\
B-122  &HR13 & 07.05.03 & 05:37:22.3 & 5400  & 1.15 & 0.96$''$  &  85 &-36.900 & -36.1 &  -15.86 & -26.9 \\
 &HR14 & 23.06.03 & 03:50:00.5 & 9900  & 1.025  & 0.68$''$       &  99 & -35.765 & & -36.87 &  \\
 &HR15 & 07.06.03 & 03:49:11.9 & 5400  & 1.15   & 0.96$''$        & 114 & -35.640 & & -27.83 &  \\
B-128  &HR13 & 07.05.03 & 05:37:22.3 & 5400  & 1.15  & 0.96$''$  &  90 &-33.331 & -32.7 & -12.28 & -23.1 \\
 &HR14 & 23.06.03 & 03:50:00.5 & 9900  & 1.025  & 0.68$''$       & 105 & -32.982 & & -33.04 &  \\
 &HR15 & 07.06.03 & 03:49:11.9 & 5400  & 1.15   & 0.96$''$        & 106 & -31.866 & & -24.04 &  \\
B-130  &HR13 & 07.05.03 & 05:37:22.3 & 5400  & 1.15  & 0.96$''$  &  78 &-36.713 & -34.9 & -15.67 & -25.3 \\
 &HR14 & 23.06.03 & 03:50:00.5 & 9900  & 1.025  & 0.68$''$       &  96 & -34.756 & & -34.82 &  \\
 &HR15 & 07.06.03 & 03:49:11.9 & 5400  & 1.15   & 0.96$''$        & 108 &-33.225 & & -25.41 &  \\
B-134  &HR13 & 07.05.03 & 05:37:22.3 & 5400  & 1.15 & 0.96$''$  & 122 & -39.272 & -37.8 & -18.22 & -28.2 \\
 &HR14 & 23.06.03 & 03:50:00.5 & 9900  & 1.025  & 0.68$''$       & 103 & -38.512 & & -38.57 &  \\
 &HR15 & 07.06.03 & 03:49:11.9 & 5400  & 1.15   & 0.96$''$        & 123 & -35.634 & & -27.81 &  \\
F-121  &HR13 & 11.06.03 & 08:46:34.2 & 8100  & 1.39  & 0.36$''$  &  92 &-29.813 & -29.2 & -24.48 & -28.0 \\
       &HR14 & 20.07.03 & 02:25:59.6 & 16500  & 1.01  & 1.03$''$ &  83 & -29.373 & & -29.37 &  \\
       &HR15 & 26.06.03 & 05:50:32.3 & 6000 & 1.06  & 0.69-0.86$''$ & 98 &-28.467 & & -30.27 &  \\
\noalign{\smallskip}
\hline
\end{tabular}
\end{flushleft}
\label{logobs}
\end{table*}

\begin{table*}
\caption[1]{
Photometric stellar parameters derived using the calibrations by Alonso et al. (1999)
 for $V-I$, $V-K$, $J-K$, bolometric corrections, bolometric magnitudes
and corresponding gravity log $g$,
and final spectroscopic parameters.   }
\begin{flushleft}
\begin{tabular}{cccccccccccccc}
\noalign{\smallskip}
\hline
\noalign{\smallskip}
\hline
\noalign{\smallskip}
& \multicolumn{5}{c}{\hbox{}} Photometric\phantom{-} parameters  & \multicolumn{5}{c}{\hbox{}}
 Spectroscopic\phantom{-} parameters\\
\cline{2-7}  \cline{9-14}    \\
{\rm star} & T($V-I$) & T($V-K$)  &  $T(J-K_{\rm TCS}$) &  ${\rm BC_{V}}$ &
${\rm M_{bol}}$ &
log g & & T$_{\rm eff}$ &
log g &[FeI/H] & [FeII/H] & [Fe/H] & ${\rm v_t }$ \\
 & K & K &  K &   & & & & K & & &  &  &  km s$^{-1}$  \\
\noalign{\smallskip}
\noalign{\hrule}
\noalign{\smallskip}
B-008 & 4867 & 4615 & 4597 & $-0.32$ & $0.93$ & 2.52 & & 4600 & 2.0 & -1.00  & -1.06 & $-1.03$ & 1.40 \\
B-107 & 4745 & 4465 & 4311 & $-0.37$ & $0.97$ & 2.50 & & 4900 & 2.1 & -1.15  & -1.06 & $-1.11$ & 1.40 \\
B-108 & 4349 & ---- & ---- & $-0.58$ & $1.50$ & 2.55 & & 4700 & 2.6 & -1.12  & -1.08 & $-1.10$ & 0.80 \\
B-118 & 4671 & 4552 & 4558 & $-0.40$ & $1.03$ & 2.49 & & 4700 & 2.6 & -0.87  & -0.81 & $-0.84$ & 1.30 \\
B-122 & 4635 & ---  & ---  & $-0.41$ & $1.04$ & 2.48 & & 4800 & 2.6 & -0.87  & -0.87 & $-0.87$ & 1.10 \\
B-128 & 4660 & 4589 & 6914 & $-0.40$ & $1.28$ & 2.58 & & 4800 & 2.7 & -0.79  & -0.79 & $-0.79$ & 1.30 \\
B-130 & 4753 & ---  & ---  & $-0.36$ & $1.28$ & 2.62 & & 4800 & 2.3 & -1.09  & -1.10 & $-1.09$ & 1.40 \\
B-134 & 5000 & 4912 & 5212 & $-0.28$ & $0.94$ & 2.57 & & --- & --- & ---  & --- & --- & --- \\
%B-134 & 5000 & 4912 & 5212 & $-0.28$ & $0.94$ & 2.57 & & 5200 & 1.1 & -1.37  & -1.28 & $-1.33$ & 0.80 \\
F-121 & 4627 & --- & --- & $-0.42$ & $1.44$ & 2.64   & & 4750 & 2.3 & -1.14  & -1.15 & $-1.15$ & 1.30 \\

\noalign{\smallskip} \hline \end{tabular}
\end{flushleft}
\label{tabteff}
\end{table*}
%===============================================================================

\subsection{Spectra}

Spectra for about  200 giants in  Baade's Window were obtained with
FLAMES+GIRAFFE  at the Very  Large Telescope, within our large program
for a spectroscopic characterization  of bulge field stars (Zoccali et
al.  2006, 2008; Lecureur  et al.   2007).  The four  fields  of this survey
include a few bulge globular clusters, and some fibres were positioned
close to their centers in order to determine consistent abundances for
bulge field and cluster stars.  Having measured the radial velocity of
the  200  spectra  in  this  field,  candidate  cluster  members  were
identified  by   selecting  stars with  radial   velocities
 and coordinates within a radius of 2.7 arcmin from the
cluster  center, therefore within the core radius.
  Nine candidates were then  identified, eight of which
turned out to  have very similar metallicity (lower  than the bulk  of
bulge stars) and  hence appeared to belong  to NGC~6522. 

\begin{figure}[ht]
\psfig{file=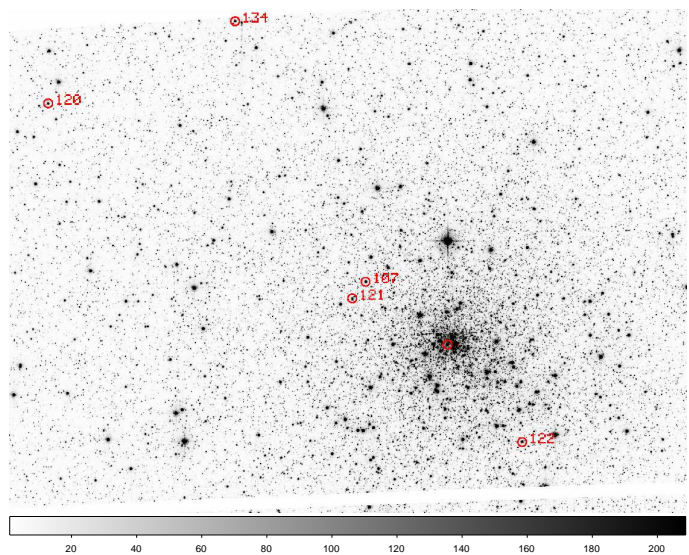,angle=0,width=9cm}
\caption {ACS image of central parts of NGC 6522.}
\label{manuds9}
\end{figure}

\begin{figure}[ht]
\centerline{
\psfig{file=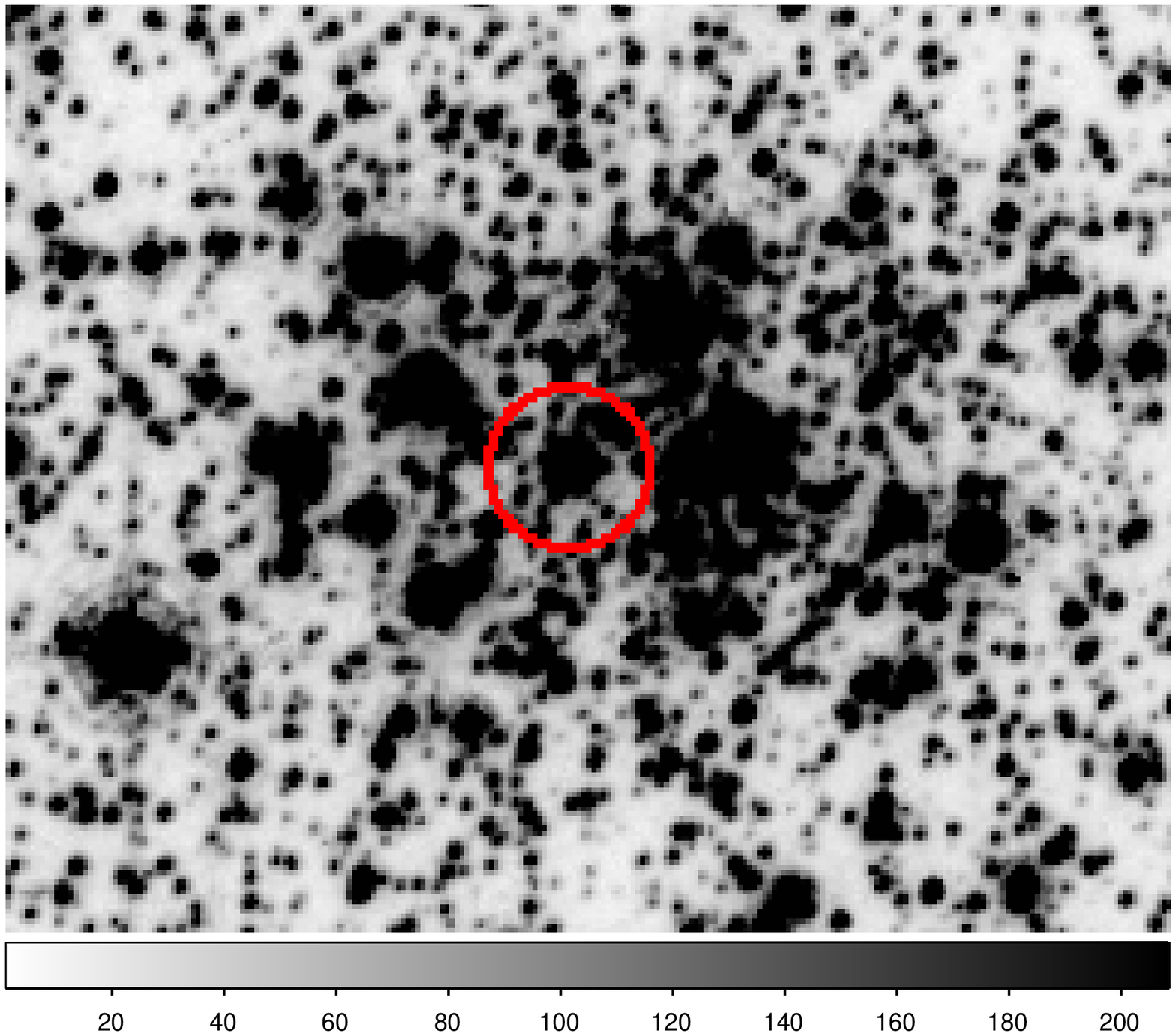,angle=0,width=4cm}
\psfig{file=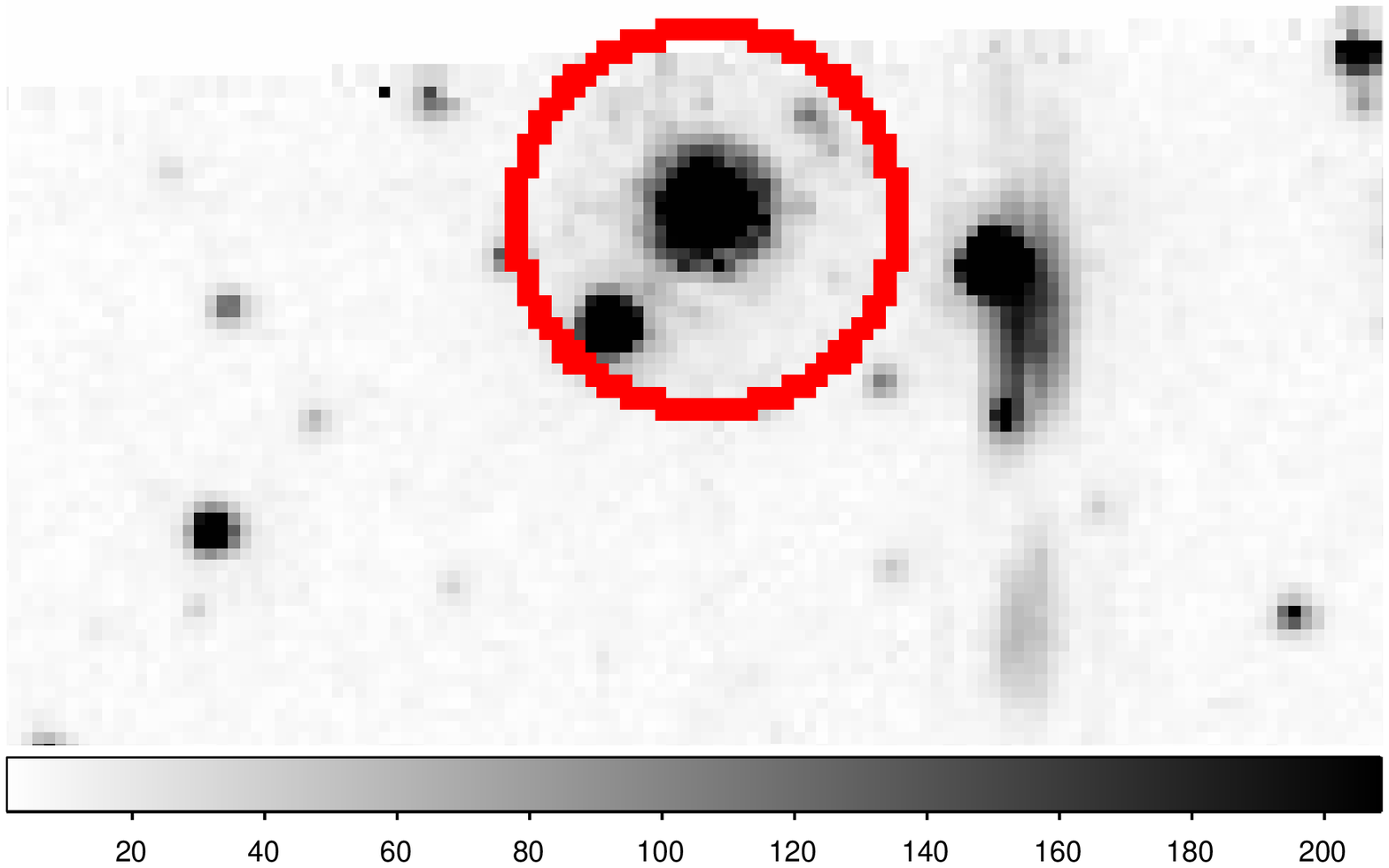,angle=0,width=4cm}
}
\caption {Zoom of ACS image in the regions of B-108 and
B-134. B-108 is located near the center of NGC 6522, and
B-134 has a fainter companion. The image of B-134 is at the
border of the ACS image.}
\label{zoom}
\end{figure}

High resolution spectra of 9 stars  in NGC 6522, in the wavelength range
$\lambda\lambda$ 6100-6860 {\rm   \AA}, were obtained 
  through the GIRAFFE setups HR13 ($\lambda\lambda$ 6120-6402 {\rm \AA}), 
 HR14 ($\lambda\lambda$ 6381-6620 {\rm \AA})
 and HR15 ($\lambda\lambda$ 6605-6859 {\rm \AA}), at a resolution
R=22,000. Log of  observations are given in  Table \ref{logobs}. 
 The   spectra   were  flatfielded,  optimally-extracted   and
wavelength calibrated    with the  GIRAFFE  Base-Line  Data  Reduction
Software  pipeline (girBLDRS\footnote{http://\-girbldrs.\-sourceforge.net}).
Spectra  extracted  from different frames were  then  summed,  and the
final spectra of $\sim 15$ fibres positioned on empty sky regions were
further combined together and subtracted from each star spectrum.
S/N ratio  were measured in the co-added spectra,
at  several wavelength regions encompassing around  2  to 3 {\rm \AA},
from 6200 to 6500 {\rm \AA}, and the values reported  in Table~\ref{logobs}
are the mean of these measurements.
  No clear continuum window could be identified in the
spectra  from the  HR15  setup, hence  S/N values were  not measured  in this
region. 
The  equivalent  widths were measured  using  the  automatic  code
DAOSPEC, developed by Stetson \& Pancino
 (2008). The stars B-108 and B-134 presented problems to converge
in terms of spectroscopic parameters, and a check
line by line was needed in order to eliminate 
 from the list any lines showing blends or cosmic ray hits.
A check on zoomed parts of the ACS image (Fig. \ref{manuds9}), given in Figs.
\ref{zoom}, shows that B-108
is near the cluster center, therefore in a very crowded region, where 
contamination could be possible, whereas B-134 has a fainter companion, having a
probable contamination of lines. B-134 was finally discarded from the sample,
due to a clear contamination of its spectrum.

A radial  velocity  v$_{\rm r} = -33.5\pm0.7$   km s$^{\rm {-1}}$  or
heliocentric radial velocity v$^{\rm hel}_{\rm r}$ = -24.9$\pm$0.7 km s$^{\rm {-1}}$ was found
for NGC 6522, in good agreement with values of 
v$^{\rm hel}_{\rm r}$ = -18.3$\pm$9.3 km s$^{\rm {-1}}$ by Rutledge et al. (1997a,b),
v$^{\rm hel}_{\rm r}$ = -28.5$\pm$6.5 km s$^{\rm {-1}}$ derived by Terndrup et al. (1998), and 
 the value  of v$^{\rm hel}_{\rm r}$ = -21.1$\pm$3.4 reported  in the compilation by Harris
(1996).

%_______________________________ STELLAR PARAMETERS___________________
%--- 3 ---
\section{Stellar Parameters}

\subsection{Temperatures}

Terndrup  \& Walker (1994) obtained BVI photometry
 on several fields of Baade's Window and NGC 6522. 
Fits to CMDs relative to
other globular clusters having the same Red Giant Branch (RGB) 
morphology, resulted in E(B-V)=0.52 for B-V=0   stars,
 corresponding to  about 0.44-0.45 for our K and M stars.
  Terndrup  et al.  (1998) derived A$_{\rm V}$=1.4  
 on  a proper motion cleaned CMD of NGC 6522.
 This gives again about  E(B-V)=0.45 for K and M stars. We
conclude that  NGC 6522 has a reddening close to 0.44-0.45, compatible
with other authors (see Table \ref{tab1}).                                                        
For the present analysis we  adopt  the extinction law given  by  Dean 
et al.   (1978) and  Rieke  \& Lebofsky (1985), namely,  
R$_{\rm V}$ = A$_{\rm V}$/E($B-V$) = 3.1,
 E($V-I$)/E($B-V$)=1.33,
 E($V-K$)/E($B-V$)=2.744,
E($J-K$)/E($B-V$)=0.527, implying in colour corrections of
A$_{\rm I}$/E($B-V$) = 1.77; A$_{\rm J}$/E($B-V$) = 0.88,
A$_{\rm H}$/E($B-V$) = 0.55;  A$_{\rm K}$/E($B-V$) = 0.356.

Effective temperatures were derived from  $V-K$, $V-I$ and $J-K$ using
the colour-temperature calibrations of Alonso et al.  (1999, hereafter
AAM99), and  using ($V-I$)$_{C}$=0.778($V-I$)$_{J}$    (Bessell 1979).
The $J,H,K_S$ magnitudes  and colours were  transformed from the 2MASS
system to  CIT (California Institute of  Technology), and from this to
TCS (Telescopio Carlos S\'anchez),  using the relations established by
Carpenter (2001)  and  Alonso et  al.  (1998). As mentioned above, 
no JHK$_s$ magnitudes  are given for B-108, B-122, B-130, F-121.

 The derived photometric effective
temperatures are  listed in  Table~\ref{tabteff}. 

\subsection{Gravities}

The classical  relation:
\[
\log g_*=4.44+4\log \frac{T_*}{T_{\odot}}+0.4(M_{\rm bol}-4.75)+\log \frac{M_*}{M_{\odot}} 
\]
was  used, adopting T$_{\odot}$=5770 K, M$_*$=0.85 M$_{\odot}$ and M$_{\rm bol \odot}$
= 4.75.

A distance modulus of ($m-$M)$_0$ = 13.91 
 together   with a  total extinction  A$_V$    = 1.72 were adopted (Barbuy et al. 1998) 
 The   bolometric  corrections from AAM99 and
corresponding gravities are given in Table~4.

%__________________________________________________
%---- 4 ----

\section{Atomic and molecular data}

The \ion{Fe}{I} line list and respective oscillator strengths
used were described in Zoccali et al. (2004), Barbuy et al. (2006a, 2007),
and reported in Table A.1, where they are compared with values given in
the NIST database (Fuhr \& Wiese 2006). 
  Five measurable \ion{Fe}{II}  lines, and their respective
oscillator strengths from Bi\'emont et al. (1991), and renormalized by
Mel\'endez \&   Barbuy (2009),   were  used to check 
whether ionization equilibrium was verified. 

In Barbuy et al. (2006a) the damping constants and gf-values 
selected in Zoccali et al. (2004) were revised concerning NaI,  MgI, SiI, CaI, TiI and
TiII lines.
The damping constants for all lines had been computed where possible, and in particular
for most  of the \ion{Fe}{I} lines,  using  the collisional broadening
theory of  Barklem  et al. (1998,  2000  and references therein).
For NaI,  MgI, SiI, CaI, TiI and TiII lines we
adopted  a  mean of $\gamma$(Barklem)/$\gamma$(best  fit)$\approx$1.5
(cf. Barbuy et al. 2006a).
The adopted oscillator strengths log gf and interaction constants C$_6$
are given in Table \ref{tablines}.
                                                             
For  the  forbidden oxygen  line  [OI]6300 {\rm   \AA}  we adopt  the
oscillator strength derived by  Allende Prieto et al. (2001):
log gf = $-9.716$.                                                                
For lines  of  the heavy  elements BaII, LaII  and  EuII,  a hyperfine
structure was taken into account,  based on the hyperfine constants
and splittings by
Lawler et al.  (2001a) for EuII 6645 {\rm \AA},  
Lawler et  al. (2001b) for  LaII 6390 {\rm \AA} and
McWilliam (1998) for BaII 6141 and constants and central wavelength
 from Rutten (1978), and hyperfine structure computed employing code
made available by A. McWilliam for BaII 6496 {\rm \AA}. Solar isotopic
ratios, and total log gf values from Hill et al. (2002), Lawler et al. (2001b)
and Rutten (1978) were adopted, as indicated in Table \ref{tablines}.
Molecular lines
of CN  (A$^2$$\Pi$-X$^2$$\Sigma$), C$_2$ Swan (A$^3$$\Pi$-X$^3$$\Pi$), TiO
(A$^3$$\Phi$-X$^3$$\Delta$) $\gamma$   and  TiO (B$^3$$\Pi$-X$^3$$\Delta$)
$\gamma$' systems  are taken   into  account.
Solar abundances were adopted from Grevesse \& Sauval (1998), except for
oxygen  where  $\epsilon$(O)  = 8.77  was  assumed, as
recommended by Allende Prieto et  al. (2001) for the  use of 1-D model
atmospheres.

%-------
%--------------------------------------------------------------------

%-------
%--------------------------------------------------------------------

%__________________________ IRON _____________________________________
%---- 5 ----

\section{Iron abundances}

Photospheric 1-D models for the sample giants  were extracted from the
new MARCS model atmospheres grid (Gustafsson   et al. 2008).
The LTE  abundance  analysis and  the spectrum  synthesis calculations
were performed using the code described  in Cayrel  et al.
(1991), Barbuy et al. (2003) and Coelho et al. (2005). An Iron abundance
of $\epsilon$(Fe)=7.50 (Grevesse \& Sauval 1998) was adopted.

The line list of \ion{Fe}{I} and \ion{Fe}{II} lines was used in the derivation
of stellar parameters, where lines with equivalent widths 
20 $<$ EW $<$ 130 m${\rm \AA}$ were not considered.
 The line list of Fe lines, together with measured
equivalent widths is given in Table A\ref{gf}.

The stellar   parameters  were  derived    by initially  adopting  the
photometric effective  temperature  and  gravity,  and   then  further
constraining  the  temperature by imposing  excitation equilibrium for
\ion{Fe}{I} lines.
% as shown in Figs.~\ref{abon2} and \ref{abon3}. A few  \ion{Fe}{II} lines
Five  \ion{Fe}{II} lines were measurable, allowing to derive gravities
imposing agreement between \ion{Fe}{I} and
\ion{Fe}{II} abundances (ionization equilibrium). Microturbulence velocities v$_t$  
were  determined by canceling the trend of \ion{Fe}{I} abundance vs.  
equivalent width, using predicted EWs, as explained  in
Zoccali et al. (2008). The analysis uses the same procedures described
in Zoccali et al. (2008) for the full sample of 800 bulge giants, 
with the differences that for NGC 6522 we use the
known distance and reddening values, and  the ionization equilibrium
between FeI and FeII to derive gravities (whereas photometric gravities
were adopted in Zoccali et al. 2008).

The final spectroscopic parameters T$_{\rm eff}$, log g, [\ion{Fe}{I}/H],
 [\ion{Fe}{II}/H],  [Fe/H] and
 v$_t$ values  are reported in  the  last columns
  of Table~\ref{tabteff} and they
 were adopted for the  derivation of abundance ratios.
It is important to note that the newly revised oscillator strenghts presented
in Mel\'endez \& Barbuy (2009), for \ion{Fe}{II} lines,
 render both the gravities and the metallicities higher
than would be derived with previous sets of log gf values.

%-------
%--------------------------------------------------------------------

%-------
\begin{figure}[ht]
\psfig{file=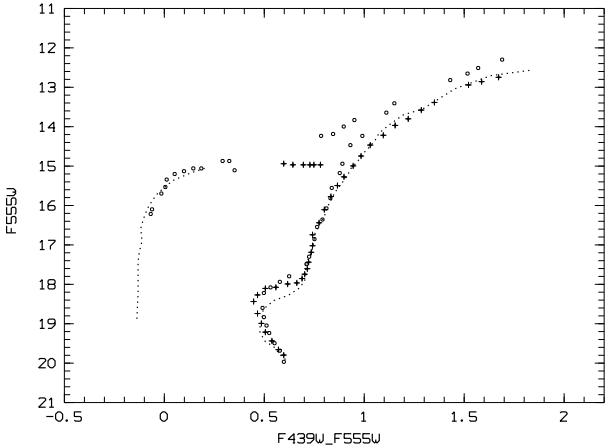,angle=-90,width=9cm}
\caption {Mean locis in V vs. V-I CMD of NGC 6522 (dots)
compared to M5 (open circles) of [Fe/H]=-1.2, and 47 Tuc (crosses)
of [Fe/H]=-0.7, based on HST-WFPC2 data from Piotto et al. (2002). }
\label{cmdcomp}
\end{figure}
%--------------------------------------------------------------------

%-------
\begin{figure}[ht]
\psfig{file=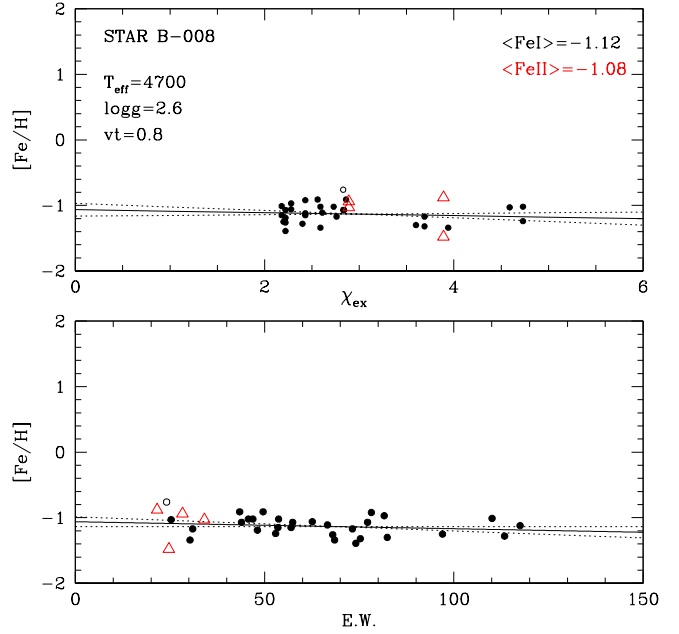,angle=0,width=9cm}
\caption {FeI and FeII abundances vs. $\chi_{\rm exc}$ (eV) and W$_{\lambda}$
(m{\rm \AA})for star NGC 6522: B-8.
% Symbols: filled dots: FeI lines; open triangles:FeII lines; open dots:
 \label{abon2}}
\end{figure}
%--------------------------------------------------------------------

\subsection{Errors}

 The errors  within the spectroscopic parameter determination are given in Table \ref{errors},
applied to the sample star NGC 6522: B-128.
The error  on the slope in the  FeI $vs.$ ionization potential implies
an error in  the temperature of $\pm$100 K   for the sample  
stars. An uncertainty of the order of  0.2 km s$^{-1}$ on the microturbulence
velocity is estimated from the imposition of constant value of [Fe/H] as
a function of EWs.
 Errors are given on FeI and FeII abundances,
  and other element abundance ratios, induced  by   a change of  $\Delta$T${\rm
eff}$=+100 K, $\Delta$log  g  =+0.2,  $\Delta$v$_{\rm  t}$ =  0.2  km
s$^{-1}$, and total error is given in the last column of Table \ref{errors}.
The errors indicated are to be added to the \ion{Fe}{I}, \ion{Fe}{II}
abundances and abundance
ratios derived in this work. It is also important to make clear
that the errors in \ion{Fe}{I} and \ion{Fe}{II} abundances 
are not propagated into the abundance ratios.

\begin{table}[ht!]
\caption{Abundance uncertainties for a $\Delta$T$_{\rm eff}$ = 100 K,
$\Delta$log g = 0.2, $\Delta$v$_{\rm t}$ = 0.2 km s$^{-1}$ and
corresponding total error. The errors are to be
added to the reported abundances. The values given are a mean
derived from the  errors  from each line.  } 
\label{errors}
\[
\begin{array}{lcccc}
\hline\hline
\noalign{\smallskip}
\hbox{Abundance} & \hbox{$\Delta$T} & \hbox{$\Delta$$\log$ g} & \hbox{$\Delta$v$_{t}$} & \hbox{($\sum$x$^{2}$)$^{1/2}$} \\
\hbox{} & \hbox{( 100 K)} & \hbox{ (+ 0.2 dex) } & \hbox{ (+ 0.2 kms$^{-1}$}) & \hbox{} \\
\hbox{(1)} & \hbox{(2)} & \hbox{(3)} & \hbox{(4)} & \hbox{(5)} \\

\noalign{\smallskip}
\hline
\noalign{\smallskip}
\noalign{\vskip 0.1cm}
\noalign{\hrule\vskip 0.1cm}
\noalign{\vskip 0.1cm}
\multicolumn{5}{c} {\bf NGC6522-B-128}\\
\noalign{\vskip 0.1cm}
\noalign{\hrule\vskip 0.1cm}
\hbox{[FeI/H]}          &  +0.06  &   0.00 & -0.08 & 0.10 \\
\hbox{[FeII/H]}         &  -0.07  &  +0.11 & -0.07 & 0.14  \\
\hbox{[O/Fe]}           &  +0.08  &  0.00 & +0.02  & 0.08  \\
\hbox{[NaI/Fe]}         &  0.00  &  +0.05 & 0.00  &  0.05  \\
\hbox{[MgI/Fe]}         &  +0.05  &  +0.01 &  0.00  &  0.05   \\
\hbox{[SiI/Fe] }        & +0.15   &  0.00 & 0.00 &  0.15  \\
\hbox{[CaI/Fe]}         &  -0.03  &  +0.04 & +0.05 &  0.07  \\
\hbox{[TiI/Fe]}         &   -0.08  &  0.00 &  +0.05 &  0.09  \\
\hbox{[TiII/Fe]}        &   0.00  &   -0.05 & -0.01 &  0.05  \\
\hbox{[BaII/Fe]}        &  -0.02  &  +0.01 & +0.15 &  0.15  \\
\hbox{[LaII/Fe]}        &   0.00  &  -0.10 & 0.00  &  0.10  \\
\hbox{[EuII/Fe]}        &   0.00  &  -0.20 & 0.00  &  0.20  \\
\noalign{\vskip 0.1cm}
\hline
\noalign{\smallskip}
\end{array}
\]
\end{table}

%_______________________ ABUNDANCE RATIOS ____________________________
% ---- 6 -----

\section{Abundance ratios}
Abundances ratios  were   obtained  by means of line-by-line  spectrum
synthesis calculations compared to the observed lines,
for the line list given in Table~\ref{tablines}. 

\begin{figure*}[ht]
\centerline{
\psfig{file=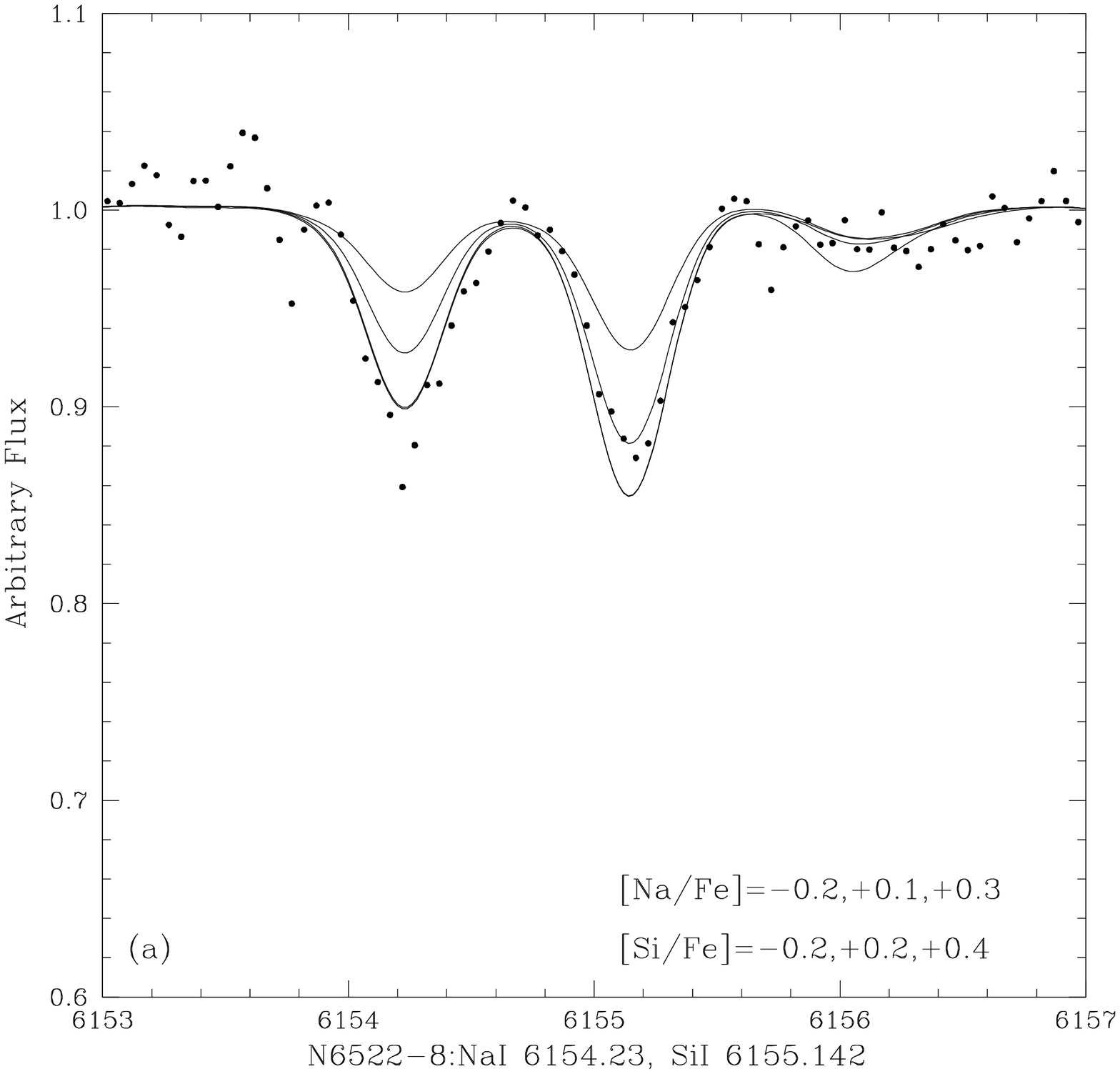,angle=0,width=5cm}
\psfig{file=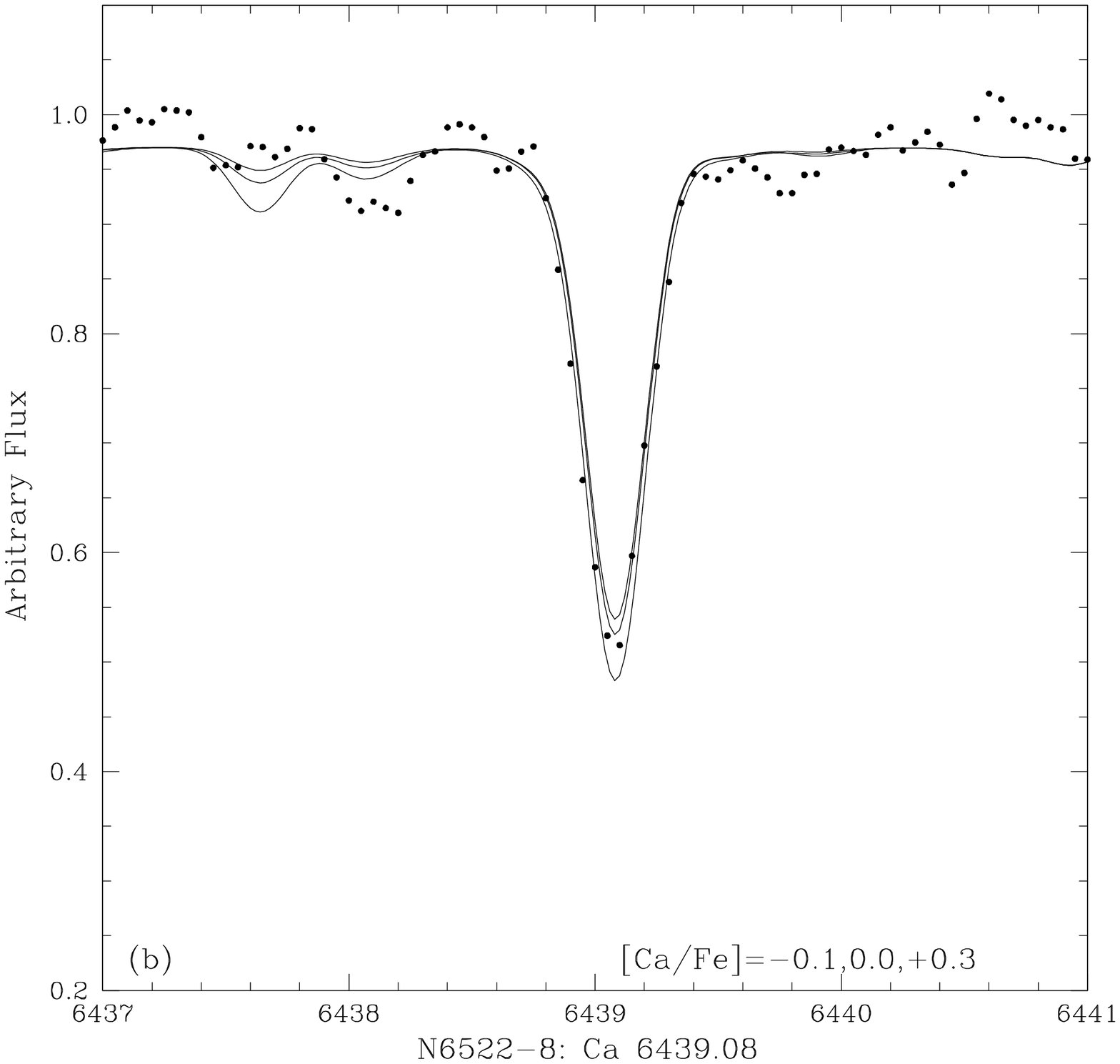,angle=0,width=5cm}
\psfig{file=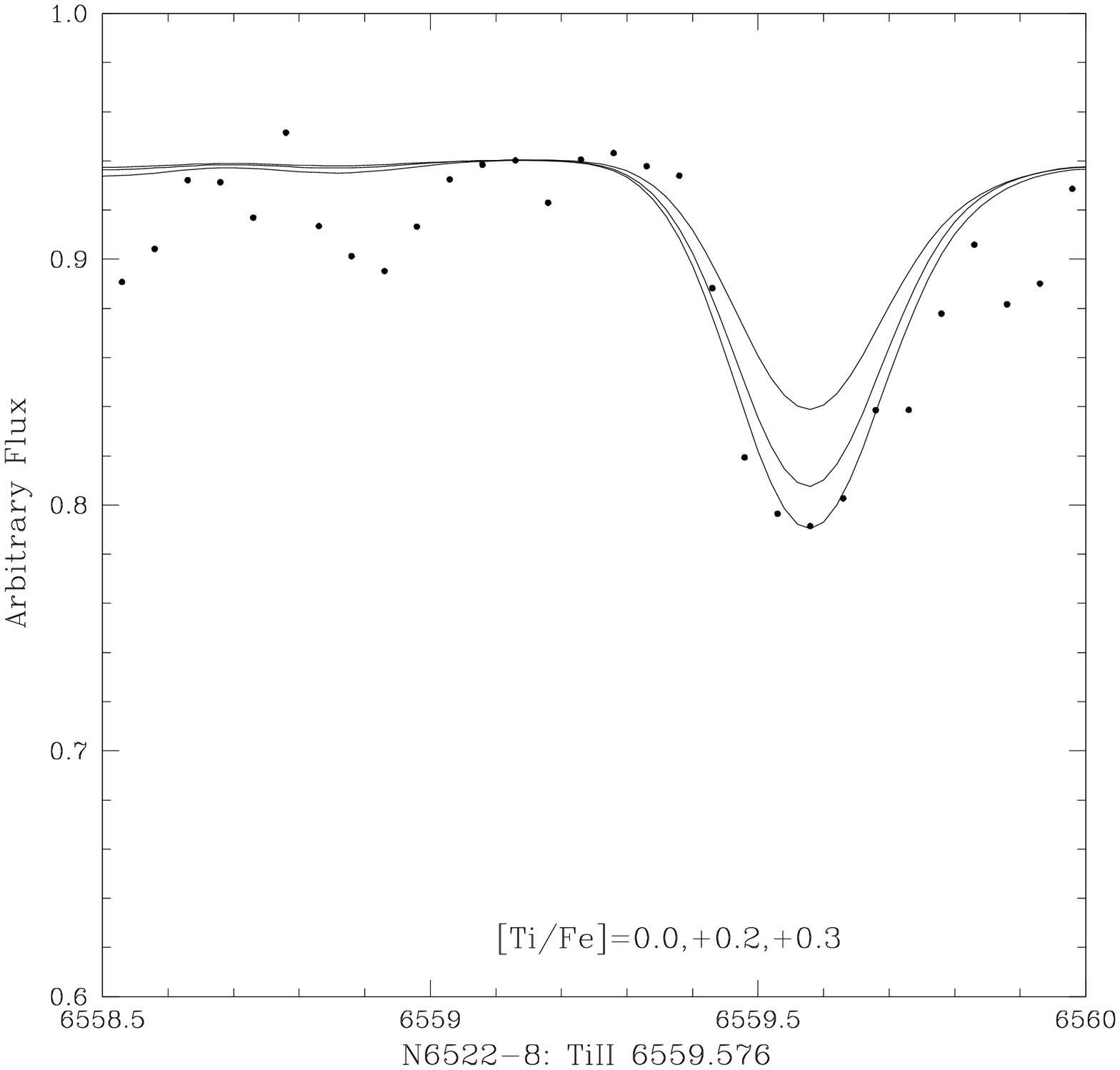,angle=0,width=5cm}}
\caption {Star B-8: (a) NaI 6154.230 and SiI 6155.142 {\rm \AA} fits.
Faint dotted line is the observed spectrum.
Solid lines show the computed spectra with abundance ratios
[Na/Fe]=-0.2,+0.1,+0.3, and [Si/Fe]=-0.2,+0.2,+0.4;
(b) Ca 6439.08 {\rm \AA} computed with [Ca/Fe]=-0.1,0.0,+0.3;
(c) TiII 6559.576 {\rm \AA} computed with [Ti/Fe]=0.0,+0.2,+0.3.
 \label{b8si}}
\end{figure*}

%The fit for the Sun, Arcturus and 
The fits to the NaI 6154.23 and SiI 6155.14 {\rm \AA} lines
in Fig. \ref{b8si}a, CaI 6439.08 {\rm \AA} in Fig. \ref{b8si}b, 
and TiII 6559.576 {\rm \AA} in Fig. \ref{b8si}c for star B-8,
 illustrate the good quality of fits.
% are  shown in Figs.~\ref{ca1}--\ref{ti} for a
%few  CaI, SiI and TiI  lines. Figs. \ref{mg} and \ref{eu} show the fits to the MgI triplet
%at 6318 and the Eu 6645  ${\rm \AA}$ line, respectively.

%-------
%--------------------------------------------------------------------

%-------
\begin{figure*}[ht]
\centerline{
\psfig{file=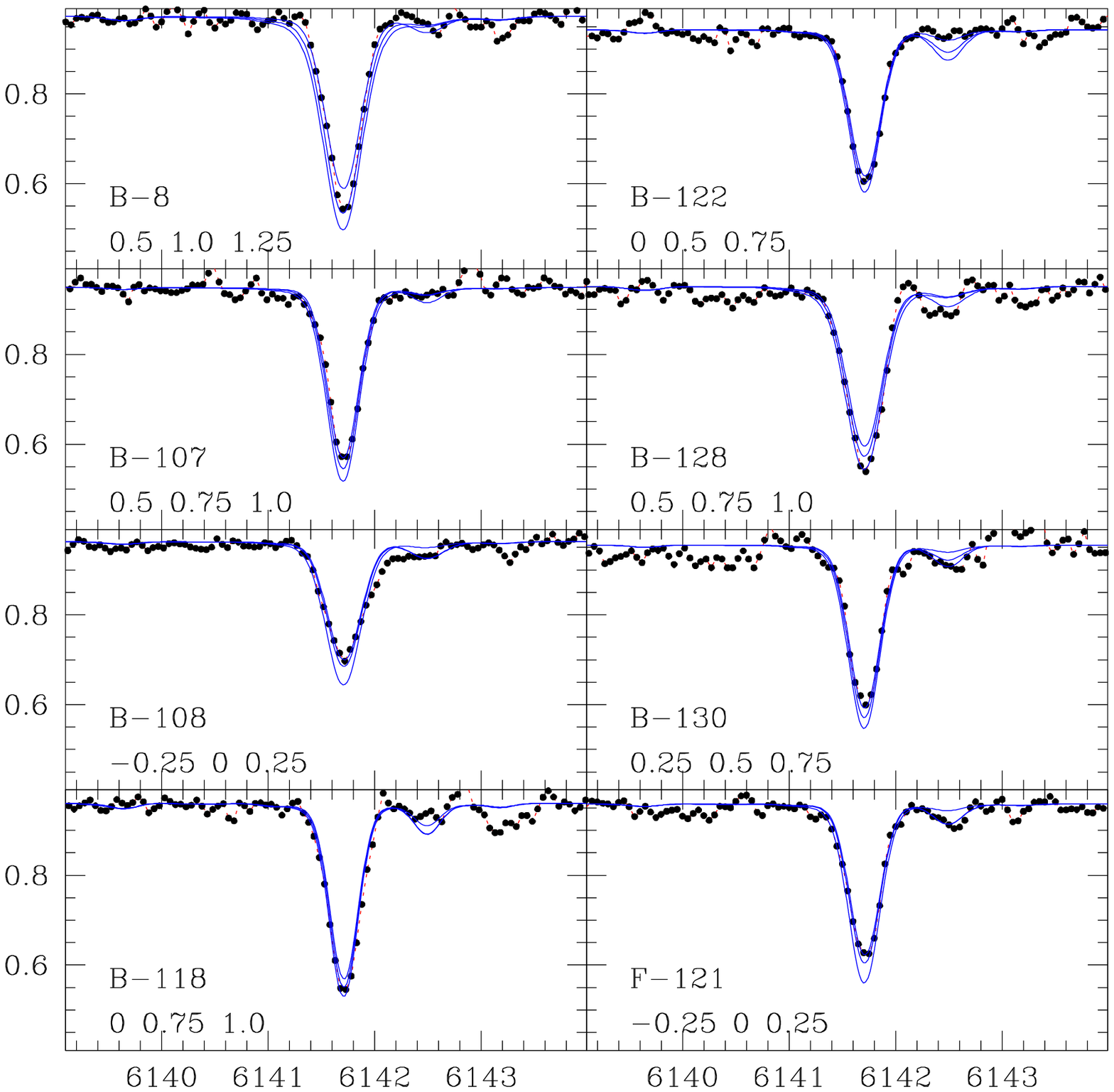,angle=0,width=18cm}}
\caption {BaII 6141.727 {\rm \AA} line in the 8 sample stars.
Observed spectra are in red and also overplotted with dots.
The [Ba/Fe] values used in each of the synthetic spectra shown
(solid lines) are indicated in the panels.
 \label{barium}}
\end{figure*}
%--------------------------------------------------------------------

The odd-Z element  sodium,
built up     during  carbon   burning,  shows a solar ratio:
[Na/Fe]$\sim$ 0.0.
The  $\alpha$-elements 
O, Mg and Si are enhanced by: [O/Fe]=+0.39, [Mg/Fe]=+0.27, [Si/Fe]=+0.25, whereas 
Ca and Ti show lower enhancements of  [Ca/Fe]=+0.17 and [Ti/Fe]=+0.16.
The r-process element Eu is enhanced by [Eu/Fe]=+0.39. The
s-elementa La and Ba are enhanced with [La/Fe]=0.35 and
[Ba/Fe]=+0.49, which is unexpected. For Ba there is also a star-to-star
variation, and these issues are discussed below.

%_____________________________________________________________________

\section{Discussion}

We  derived  a metallicity of [Fe/H]=-1.0$\pm$0.2,  and
the element abundances reported in Tables \ref{tablines} and
\ref{final},  for 8 giants in the  globular cluster NGC 6522.

\subsection{Bulge globular clusters}
 
In order to better characterize NGC 6522,
in Table \ref{globsmet} we report globular clusters
within 6$^{\circ}$$\times$6$^{\circ}$ of the Galactic center,
 classified by metallicity, and
HB morphology. We identify a group of BHB clusters combined to a moderate
metallicity of [Fe/H]$\approx$-1.0. As discussed in Barbuy et al.
(2006a, 2007) regarding HP~1 and NGC~6558, these clusters should
be very old (Lee et al. 1994).
In this inner sample there are 6 over 19 objects
 in this class of globular clusters,
and they constitute therefore a  fraction around 30\%.
The last item in  Table \ref{globsmet} concerns globular clusters
that show similar properties to NGC 6522, i.e., a BHB
and having moderate metallicity, that are located between
6$^{\circ}$ and 12$^{\circ}$ around the Galactic center; 
in this ring this class of
clusters corresponds to a lower
fraction of 16\% of clusters, considerably lower with respect to the
 30\% fraction in the inner 6$^{\circ}$$\times$6$^{\circ}$ region, 
and we thus infer that this population appears to be
more concentrated towards the Galaxy center.

As concerns the presence of RR Lyrae in the metal-poor inner
bulge globular clusters, in particular those
studied by means of spectroscopy so far (HP~1, NGC 6522, NGC 6558 and
Terzan 4), 
from the most recent RR Lyrae compilation sources (Suntzeff et al. 1991,
and the on-line catalogue published by Clement et
al. (2001)\footnote{http://www.astro.utoronto.ca/~cclement/read.html},
the following can be extracted: there are no identified RR Lyrae
stars in very metal-poor clusters Terzan 4, Terzan 9, 
and the moderately metal-poor NGC 6540. No studies on the recently discovered AL3
(Ortolani et al. 2006) are available.
For HP1 there are 15 variable stars reported by Terzan (1964a,b, 1965, 1966)
 but none has been indicated as RR Lyrae.
NGC 6558 has 9 confirmed RR Lyrae stars (Hazen 1996).

The field of NGC 6522 is very rich in RR Lyrae stars and they appear to be
enhanced towards the center of the cluster.
Seven variables have been detected within 2' from the center of NGC 6522,
but their membership is uncertain (Walker \& Mack 1986; Clement et al.
1991). However Walker and Terndrup (1991), from radial velocities and
metallicities, concluded that 4 of them should be RR Lyrae stars members of
the cluster. From the DS index they obtained an average of [Fe/H]=-1.0,
very near to the mean of the other Baade Window RR Lyrae stars they
studied with the same method, but it is considered that
 the membership of these RR Lyrae stars 
to NGC 6522 remains an open issue.

As a conclusion, we suggest that the [Fe/H]$\approx$-1.0 RR Lyrae
and our sample clusters of [Fe/H]$\approx$-1.0 and BHB, such as NGC 6522, 
could belong to the same stellar population.

\subsection{Age of NGC 6522}

In Fig. \ref{cmdcomp} we show the mean locus of NGC 6522 from
Piotto et al. (2002) using Hubble Space Telescope F439W and
F555W bands of the WFPC2 camera. Overplotted are the mean loci
of M5 (NGC 5904) of [Fe/H]=-1.2 (e.g. Yong et al. 2008) and
47 Tuc of [Fe/H]=-0.7 (Alves-Brito et al. 2005),
or -0.76 (Koch \& McWilliam 2008). The RGB
of NGC 6522 is very close to that of 47 Tuc, the latter
slightly  more metal-rich, and M5 is steeper. This confirms
the metallicity higher than usually assigned to NGC 6522
(see Table 1).
 These high metallicities combined to a blue HB are not
expected, and might have led to the lower metallicity estimates
in the past.

Fig. \ref{cmdcomp} shows  that the turnoff of  NGC 6522  appears to be about
0.2 magnitudes fainter than those of 47 Tuc and M5, when the
HBs are superimposed. This indicates an age about 2 Gyr older
for NGC 6522. The older age is in agreement with the HB morphology
because it is very blue for its relatively high metallicity.

\begin{figure}[ht]
\psfig{file=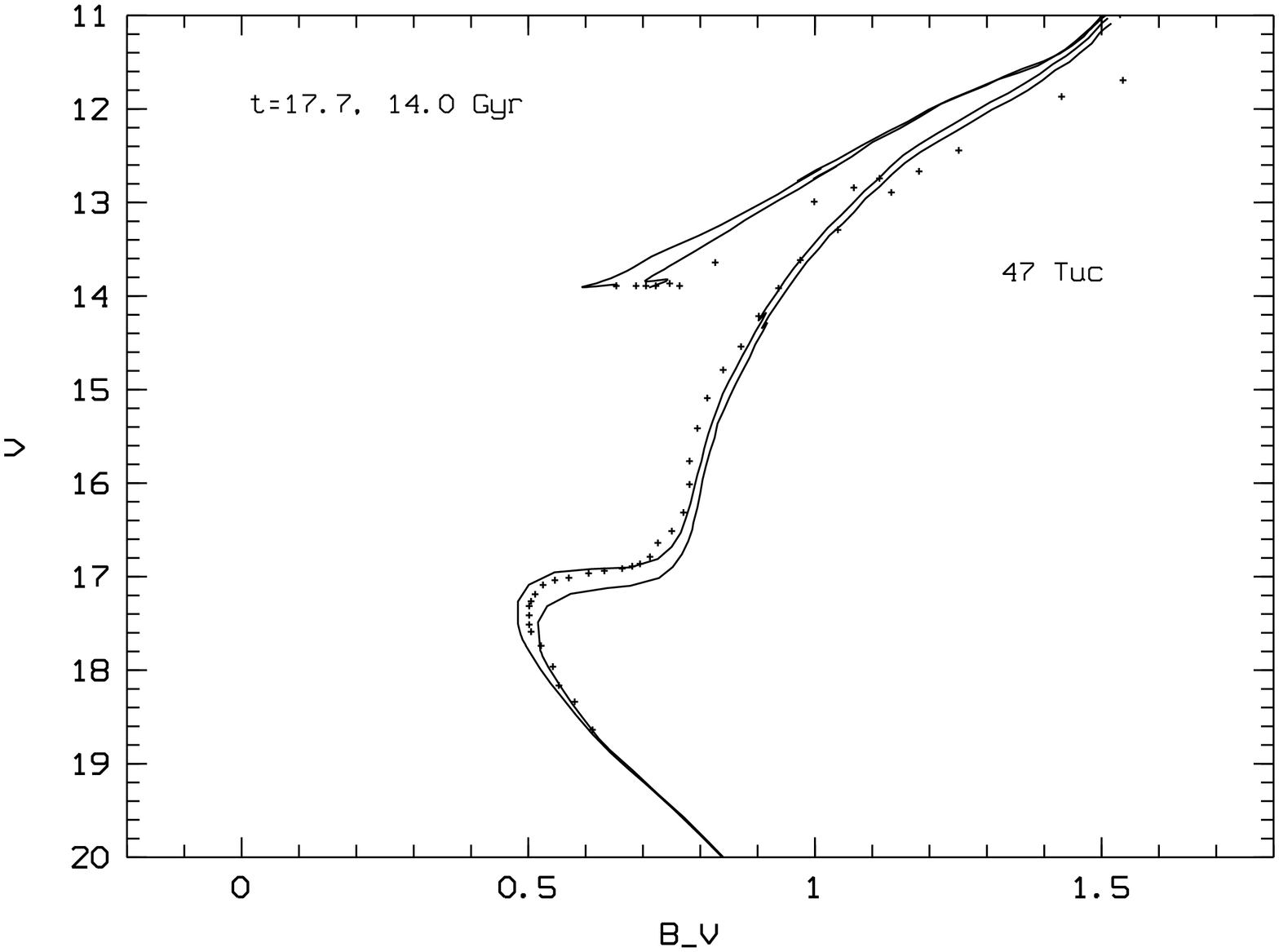,angle=-90,width=9cm}
\psfig{file=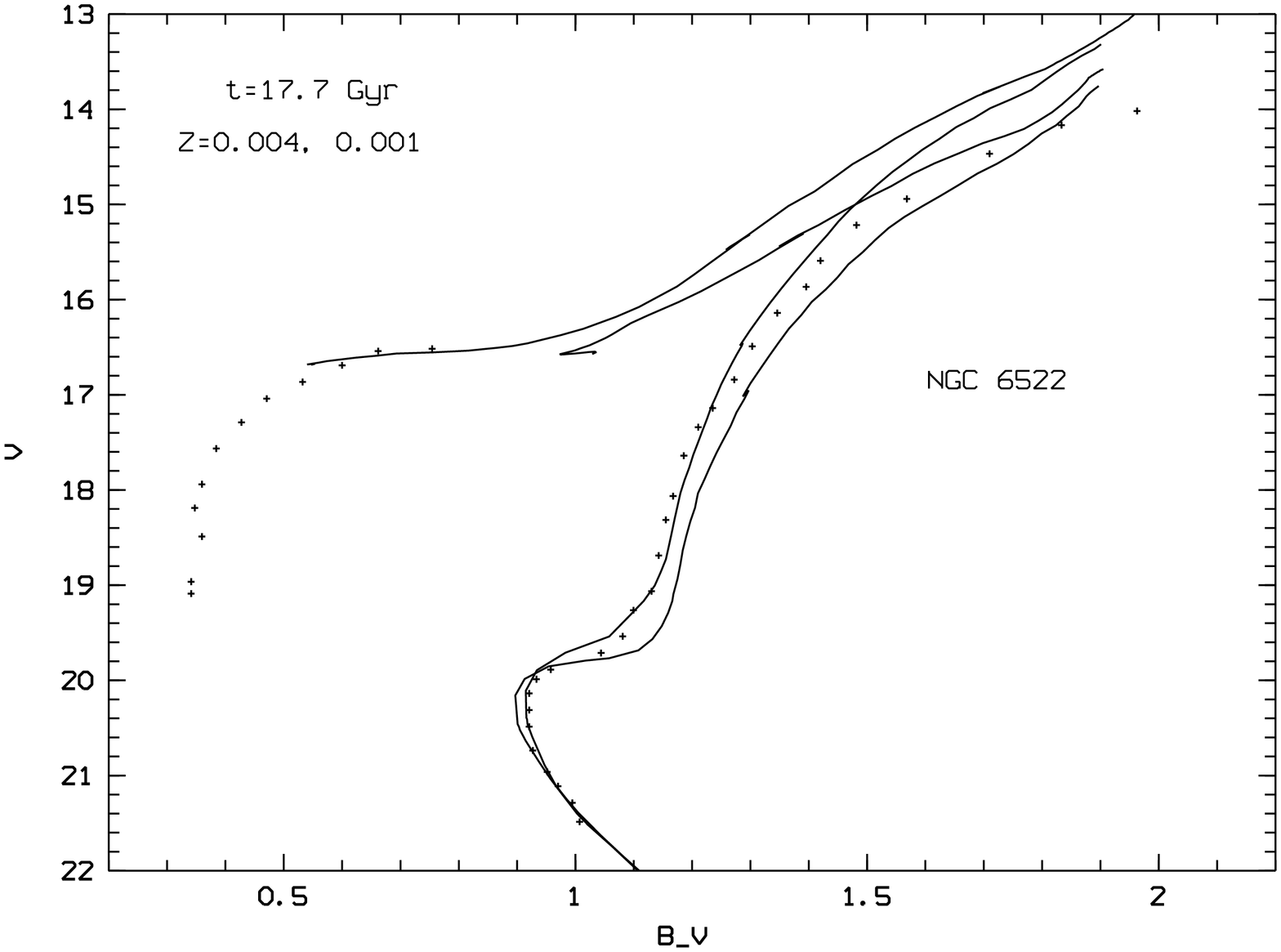,angle=-90,width=9cm}
\caption {Mean loci CMDs of 47 Tuc and NGC 6522, based on HST data from
Piotto et al. (2002), overplotted with isochrones from Girardi et al. (2000);
In the upper panel, 47 Tuc is overplotted with isochrones of Z=0.004 and ages 14.0 and 17.7 Gyr,
in the lower panel, NGC 6522 is overplotted with isochrones of age 17.7 Gyr and metallicities
of Z=0.001 and 0.004.}
\label{isoch}
\end{figure}

\begin{figure}[ht]
\psfig{file=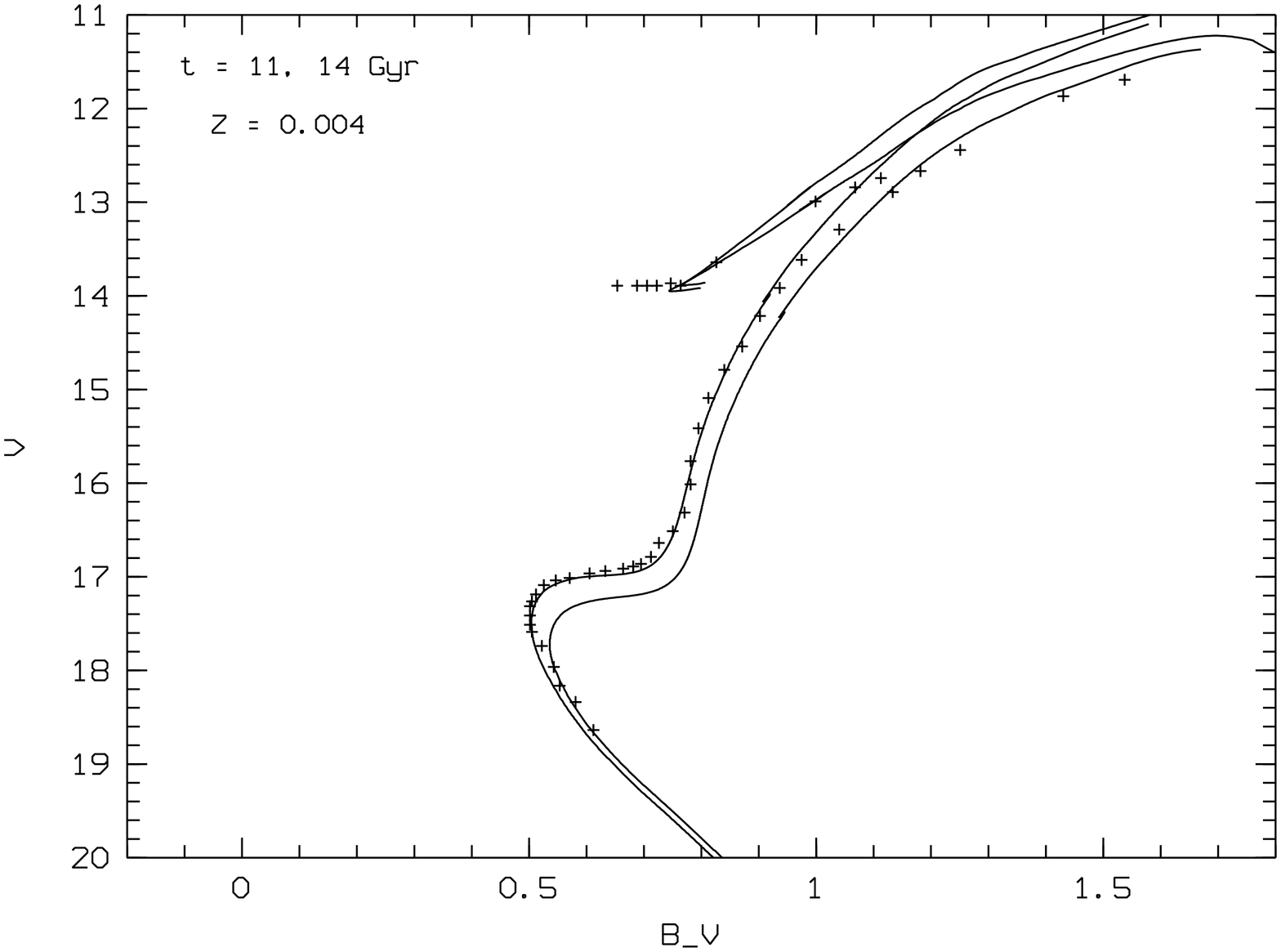,angle=-90,width=9cm}
\psfig{file=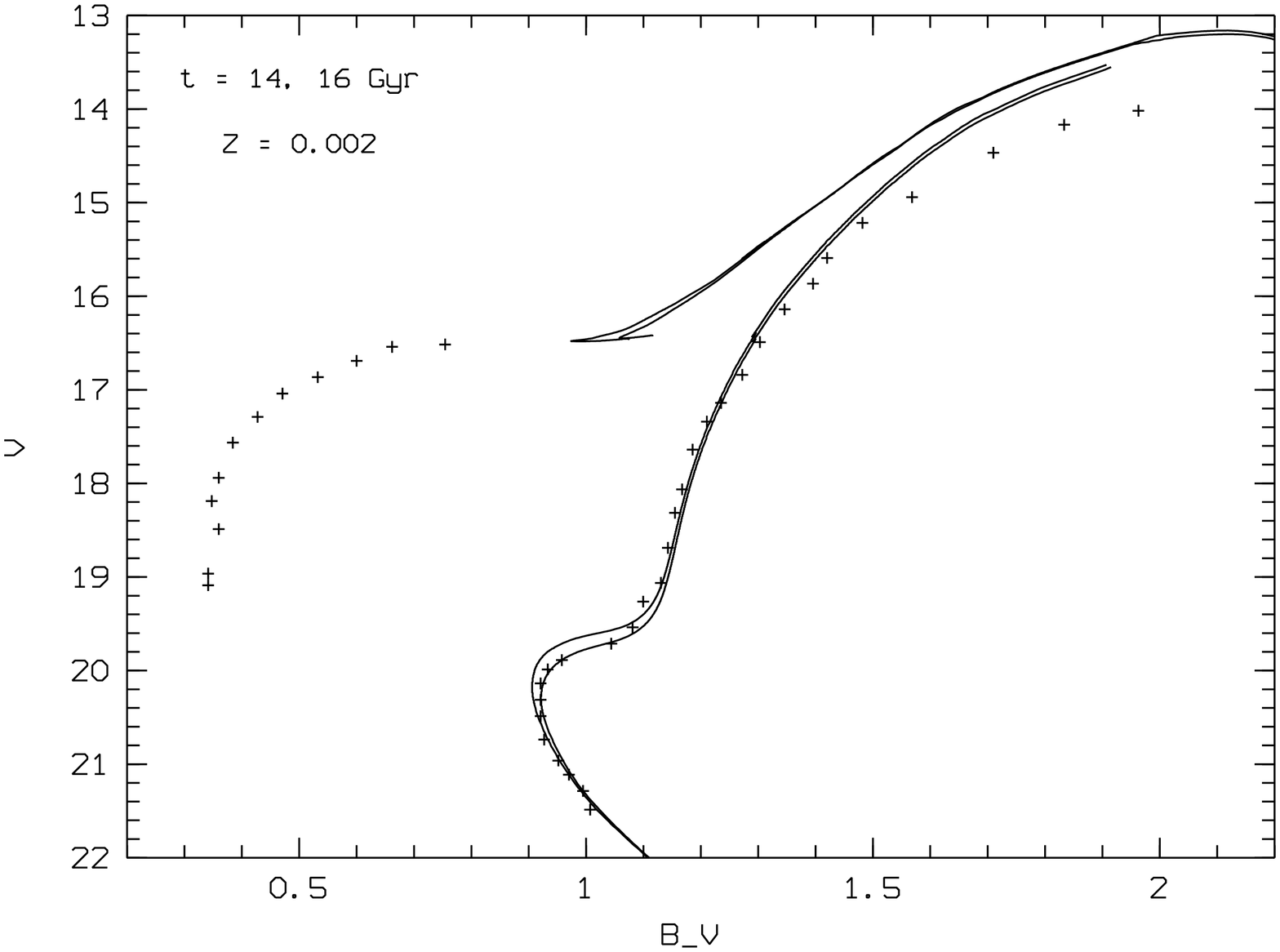,angle=-90,width=9cm}
\caption {Mean loci CMDs of 47 Tuc and NGC 6522, based on HST data from
Piotto et al. (2002), overplotted with isochrones from Pietrinferni et al. (2004);
In the upper panel, 47 Tuc is overplotted with isochrones of Z=0.004 and ages 11.0 and 14.0 Gyr,
in the lower panel, NGC 6522 is overplotted with isochrones of ages 14.0 and 16.0 Gyr
 and a metallicity of Z=0.002.}
\label{teramo}
\end{figure}

 In order to check this age difference between 47 Tuc and NGC 6522, 
in Fig. \ref{isoch} we collected data from Piotto et al. (2002) for
these two clusters, with colours transformed to B and V, and 
derived mean loci. Isochrones from Girardi et al. (2000) are then
overplotted to these mean loci CMDs. For 47 Tuc, a metallicity of
Z=0.004 ([Fe/H]~-0.7) is adopted, with ages of 14.0 and 17.7 Gyr.
It is clear that the 14 Gyr isochrone fits the data whereas this
is not the case with 17.7 Gyr;
for NGC 6522 the oldest isochrones of 17.7 Gyr for metallicities of
Z=0.004 and 0.001 are overplotted. It is clear that this very old age
fits the observed CMD with both metallicities.
We also used $\alpha$-enhanced Teramo isochrones
from Pietrinferni et al. (2004), as described in
the BASTI 2006 grid of models\footnote{http://193.204.1.602/index.html},
shown in Figs. \ref{teramo}.
 Adopting for 47 Tuc a metallicity of
Z=0.004 ([Fe/H]~-0.7), and ages of 11.0 and 14.0 Gyr, where the
11.0 Gyr one fits its CMD; and
for NGC 6522  isochrones of 14.0 and 16.0 Gyr for a metallicity of
Z=0.002 are overplotted. In this case the  old age of 14.0 Gyr
fits the observed CMD.
This makes evident the old age of NGC 6522, although of course 
it cannot be older than 13.7 Gyr (Spergel et al. 2003), but the
important result is its relative older age as compared with 47 Tuc.
 
%-------     
%-------------------------------------------------------------------- 

%-------     
%-------------------------------------------------------------------- 

Still, another way to estimate the age, despite more imprecise, is
 the HB parameter by Lee (1992) (B-R)/(B+V+R),
 where B = blue HB stars, R = red HB stars, and V = RR Lyrae.
Based on a HST CMD from Piotto et al. (2002), we have measured
25 BHB stars,  and none in the gap and red HB, assuming the standard
HB gap to be at 0.15 $<$ B-B $<$ 0.45. Therefore NGC 6522 has
an HB parameter of (B-R)/(B+V+R)=1.0, a value probably more
accurate than the value of 0.71 reported by Terndrup \& Walker (1994).
 This HB morphology together with the [Fe/H]=-1.0 metallicity
plotted in [Fe/H] vs. HB type shown for example in Lee (1992, Fig. 5)
would lead to an age at least 2 Gyr older than typical halo
clusters. The same applies to metallicities of [Fe/H]=-1.1,
and the higher the metallicity the higher the age.

In the following discussion, we argue that these clusters
might be the earliest objects in the Galaxy.

The relative ages of 64 Galactic globular clusters, resulting
from the corresponding HST Treasury program based on an ACS survey
is presented by Mar\'{\i}n-Franch et al. (2009). Their Fig. 13 shows
an important new result, which is that globular clusters
with metallicities [Fe/H] $\simgreat$ -1.4 are found in two
types in terms of age. Those of galactocentric radius
R$_{\rm GC}$ $<$ 10 kpc, are very old. According
to Mar\'{\i}n-Franch et al. this population is consistent with
a galaxy formation scenario of a rapid collapse, and the
formation of the old group of clusters within the bulge and
halo in a timescale $\simless$ 0.8 Gyr. Another possibility
is that they could also have formed before reionization, in dwarf
galaxies that later merged to form the Milky Way. The other group shows
younger ages, and are probably formed in satellite dwarf galaxies
accreted by the Milky Way.

\begin{figure}[ht]
\psfig{file=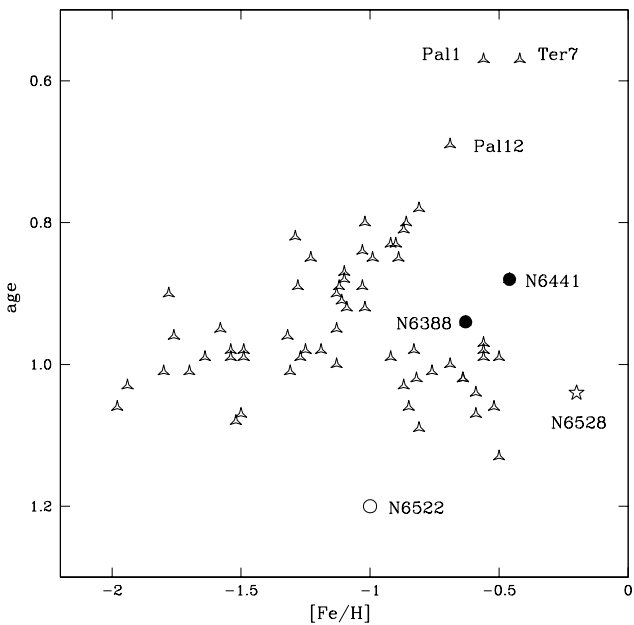,angle=0,width=9cm}
\caption {Age vs. metallicities, given in the CG metallicity
scale, of the 64 globular
clusters from Mar\'{\i}n-Franch et al. (2009), and the location
of NGC 6522 in the plot, showing it to be the oldest cluster
of this sample (open circle). The most metal-rich bulge cluster 
NGC 6528 ([Fe/H]=-0.2) is identified with an open star.
The metal-rich [Fe/H]=-0.6 clusters that have an
extended blue HB and a red HB, NGC 6388 and NGC 6441,
 are indicated by full circles. Some of the youngest clusters Pal 1,
Pal 12 and Ter 7 are also identified.}
\label{relatage}
\end{figure}

In Fig. \ref{relatage} the age and metallicity
of NGC 6522 is plotted together with the data on 64 globulars
by Mar\'{\i}n-Franch et al. (2009).
The age of NGC 6522 is assumed to be 2 Gyr older than the mean
of globulars as indicated from the Lee (1992) HB type parameter,
and the low turn-off. We adopted the metallicity scale of 
Carretta \& Gratton (1997, hereafter CG) rather than
the Zinn \& West (1984, hereafter ZW) one, for
the main reason that CG derived metallicities from clean FeI lines, 
as in the present analysis.  Among the ages reported in 
Mar\'{\i}n-Franch et al. (2009), we adopted those based on the
stellar evolution libraries by Dotter et al. (2007) with
the option of metallicity compatible with the CG scale.
 Fig. \ref{relatage} shows  that NGC 6522
 would be the oldest cluster in the sample.

We have also plotted in Fig. \ref{relatage} an estimate of relative
age for NGC 6528, that was found to be the most metal-rich
cluster in the Galaxy (Zoccali et al. 2004; Ortolani et al. 2007).
A comparison of an ACS CMD for NGC 6528 (Brown et al.
2005) of [Fe/H]=-0.2 (Zoccali et al. 2004), with that of NGC 6366 of 
[Fe/H]=-0.44 in ZW scale, or [Fe/H]=-0.59 in CG scale (Mar\'{\i}n-Franch
et al. 2009), and both show $\Delta$V(TO-HB)=3.8, giving a relative
age of 1.04 (assuming that the metallicity difference between
NGC 6528 and NGC 6366 is small enough to be neglected).
We also identify in this Fig.
 the clusters NGC 6388 and NGC 6441 (Rich et al. 1997), known to
have BHBs and high metallicity [Fe/H]$\sim$-0.6.
 These clusters, indicated by full circles in Fig. \ref{relatage},
 are different from NGC 6522 in the sense that
 NGC 6388 and NGC 6441 have both a blue and a populous red HB.

\begin{table}
\caption[1]{
Classification of inner bulge globular clusters in terms of metallicity
and HB morphology (R = Red, B = Blue). References: 1 Ortolani et al. 1997a;
2 Barbuy et al. 1998; 3 Origlia \& Rich 2004; 4 Ortolani et al. 1999a;
5 Ortolani et al. 1997b; 6 Barbuy et al. 2006a; 7 Piotto et al. 2002;
8 Rich et al. 1998; 9 Barbuy et al. 2007; 10 Ortolani et al. 2006; 
11 Ortolani et al. 1997c; 12 Bica et al. 1994; 13 Ortolani et al. 1997d;
14 Barbuy et al. 1997; 15 Origlia et al. 2005;  
16 Ortolani et al. 1999b; 17 Idiart et al. 2002; 18 Lee et al. 2004;
19 Zoccali et al. 2004; 20 Ortolani et al. 2007; 21 Alves-Brito et al. 2006;
22 Ortolani et al. 1996; 23 Origlia et al. 2002; 24 Ortolani et al. 2003;
25 Harris 1996; 26 Barbuy et al. 2006b; 27 Present work; * Note that Davidge (2000)
obtained [Fe/H]=-2.2 for Djorgovski 1. }

\begin{flushleft}
\begin{tabular}{l@{}l@{}l@{}l@{}l@{}l@{}l@{}rrr}
\noalign{\smallskip}
\hline
\noalign{\smallskip}
\hline
\noalign{\smallskip}
cluster & E(B-V) & \phantom{-}d$_{\odot}$ &
[Fe/H] &
\phantom{-}v$_r$  &
HB &
$\phantom{--}$ref.  \\
 &  & (kpc) &  &
(km.s$^{-1}$) &
 &  \\
\noalign{\vskip 0.2cm}
\noalign{\hrule\vskip 0.2cm}
\noalign{\vskip 0.2cm}
\noalign{\bf Metal-poor}
Terzan 4 & 1.8 & 8.3 & -1.6 & -50 & B & 1,2,3 \\
Terzan 9 & 1.95    & 4.9 & -1.0 & --- & B & 4  \\
\noalign{\bf Moderately metal-poor, BHB}
HP~1     & 1.21        & 6.4  & -1.0 & 46 & B & 5,6  \\
NGC 6522 & 0.45        & 7.4 & -0.86 & -25 & B & 7,27 \\
NGC 6558 & 0.38        & 7.7 & -0.97 & -197  & B  & 8,9 \\
Al~3     &0.36         & 6    & -1.3 & -- & B &  10 \\
Terzan 10 & 2.4       &  4.8    &  -1.0    & ---   & B & 2,11 \\
NGC 6540 &0.60 & 3.5 &-1.0 & --- & B & 12 \\
\noalign{\bf Moderately metal-poor, RHB}
Terzan 2 & 1.6 & 6.6 & -0.5 & --- & R & 2,13 \\
Terzan 6 & 2.3 & 5.4 & -0.5 & --- & R & 2,14 \\
UKS 1 &3.6 &10 & -0.78 &+57 & R & 11,15 \\
ESO456-SC38 & 0.90 &5.5 &-0.5 &--- & R & 2,11 \\
Terzan 1 & 2.5 & 5.2 &-1.3 &114 & R & 16,17 \\
Palomar 6 &1.36 & 6.40 &-1.0 & 180 & R & 18 \\
Djorgovski 1 &1.70  &5.6 &-0.4 &--- &R & 2,* \\
\noalign{\bf Metal-rich}
NGC 6528 &0.46 & 7.7 & -0.1 &212 & R & 19,20 \\
NGC 6553 &0.7 & 5.1 &-0.2 &-2 & R & 2,21 \\
Terzan 5 &2.39 &5.6 &-0.3 &-93 & R & 2,3,22 \\
Liller 1 & 3.7 & 6 & -0.3 &+64 & R & 20,23 \\
\noalign{\bf Moderately metal-poor, BHB, located in the ring  6$^{\circ}$ - 12$^{\circ}$}
NGC 6325 & 0.89 &   6.9  &   -1.17 & 29.8  & B & 7,24,25 \\
NGC 6355 & 0.75 &   8.8  &  -1.3   & -176.9  &  B  & 7,24,25 \\
NGC 6453 & 0.61 &   11.2 &   -1.53 & -83.7  & B & 7,25 \\
NGC 6626 & 0.40 &   5.7  &  -1.0   & 17.0  &     B & 25 \\
NGC 6642 & 0.41 &   7.2  &  -1.3   & -57.2  &  B  &  7,26  \\
%NGC 6284 &  &   14.7 &  -1.32  &   &  B    & 7,22 \\
\noalign{\smallskip} \hline \end{tabular}
\end{flushleft}
\label{globsmet}
\end{table}

Another issue concerns the helium abundance:
Recio-Blanco et al. (2006) discussed the HB characteristics
of 54 BHB globular clusters, and concluded that the blue HB extension
depends on metallicity, mass of the cluster (interpreted  as
self pollution of helium) and age. More parameters could be also involved,
as for example concentration and consequent higher collisional probability
among member stars.
We believe that the hypothesis of a high helium abundance does not apply to NGC 6522:
some globular clusters include helium enriched sub-populations candidates
to explain HB blue extensions (e.g. Piotto et al. 2007, and references
therein), however,  such helium -enriched populations have
been inferred in  massive GCs, with mass in excess of $\sim
10^6\,M_\odot$, involving few of the brightest clusters
such as $\omega$ Centauri of M$_{\rm V}$ = -10.29 (Harris 1996),
which is not the case of NGC 6522 (see also Sect. 1).
% MV wcen and 2808

\subsection{Chemical enrichment scenarios}
Our results show  enhancements of 
[O/Fe]=+0.4, [Mg/Fe]=[Si/Fe]$\approx$+0.25,    
 and lower [Ca/Fe]=[Ti/Fe]$\approx$+0.15.
% Differences in the enhancements of different $\alpha$-elements were
 %also found for example by McWilliam \& Rich (1994, 2003)
% and Pomp\'eia et al. (2003).
The odd-Z element Na shows a solar ratio  [Na/Fe]$\sim$0.0.
The $\alpha$-element enhancements in 
Oxygen, Magnesium and Silicon,  together with that of 
the r-process element Eu [Eu/Fe]$\approx$+0.4 are indicative of
 a fast early enrichment by Supernovae type II.
%Also, the present  $\alpha$-enhancements are compatible with
%results for the bulge  field  stars found  by McWilliam \& Rich
%(2003). On the other hand, the solar ratios of Mg, Ca and Ti 
%may be indicative of important contribution by SN Ia type.
%Therefore in terms of formation of globulars in the bulge,
%the results are not conclusive.

The s-elements Ba and La show high ratios [La/Fe]$\approx$+0.35 and
 [Ba/Fe]$\approx$+0.5.
 An enhanced Barium abundance is also found in relatively
metal-rich globular clusters such as 47 Tucanae (Alves-Brito et al. 2005),
M71 (Ram\'{\i}rez \& Cohen 2002), and M4 (Ivans et al. 1999).
The large abundances of Ba and La, and their large star-to-star variations
are illustrated in Fig. \ref{barium}, showing the BaII 6141.727 {\rm \AA}
in the 8 sample stars. The high La and Ba abundances are consistent 
with each other, but the [Ba/Eu]=+0.1 ratio, that could give a 
measure of the s- to r- process nucleosynthesis, is too high, and clearly
the La and Ba excesses cannot be attributed to the r-process.
The s-elements excesses could be due to an s-process enrichment
of the primordial matter from which the cluster formed, or else
s-process the occurred in nearby Asymptotic Giant Branch (AGB) stars
during He shell flash episodes, and the ejected material would be
accreted by the sample stars during their formation process. This
latter explanation would also account for the large spread in the 
Ba abundances.

The star-to-star variation of the r-process element Eu are more difficult
to explain, since r-elements are only produced in supernovae events. The 
only explanation would be a primordial enrichment by different supernovae,
and no gas mixing prior to the formation of the sample stars. The
spectral region of the EuII 6645.127 {\rm \AA} line is however rather
noisy, and we prefer not to draw conclusions on these Eu abundance
variations.

\begin{figure}[ht]
\psfig{file=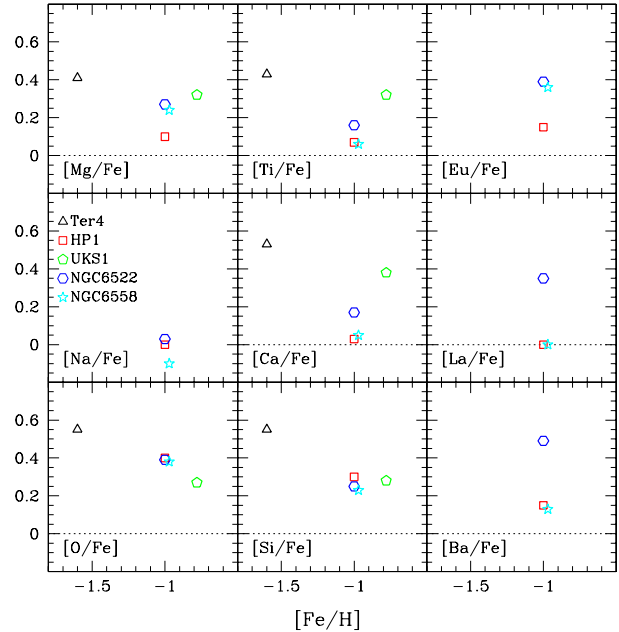,angle=0,width=9cm}
\caption {Abundance pattern of NGC 6522 compared to other
metal-poor globular clusters located in the Galactic bulge.}
\label{pattern}
\end{figure}

Finally, Table \ref{tabalpha} and  Fig. \ref{pattern} show
a comparison of abundance ratios in NGC 6522, 
 with a) results by Origlia \& Rich (2004) 
for the metal-poor cluster Terzan 4 ([Fe/H]=-1.6),
also located in the inner bulge for
which significant enhancements of  $\alpha$-elements were found,
and UKS 1, a moderate metallicity cluster ([Fe/H]=-0.78) with
moderate enhancements, 
b) the clusters  HP 1 and NGC 6558.
HP 1 and NGC 6558, similarly to NGC 6522, show a peculiar 
pattern and might be revealing characteristics 
of the early bulge chemical enrichment.
It would be of great interest to have
further analyses of individual stars in the three clusters HP-1,
NGC 6558 and NGC 6522 and other  metal-poor bulge globular clusters.

%-------
%--------------------------------------------------------------------

%-------
%--------------------------------------------------------------------

%

\section{Conclusions}

Lee (1992) pointed out that RR Lyrae in the Galactic bulge
have a peak metallicity of [Fe/H]$\approx$-1.0. This population
should be older than the halo, because being more metal-rich
these stars should be more massive, and expected to populate the red HB,
 whereas given that they populate the RR Lyrae gap, then a lower mass is required
for them, consequently implying older ages.

The metallicity of [Fe/H]$\approx$-1.0
was derived for NGC 6522, based on high resolution spectra,
 similarly to results found for HP~1 and NGC~6558. 
This relatively high metallicity was unexpected given
that in the literature, the metallicity values 
quoted are in the range [Fe/H] = -1.3 to -1.5 along the
last decades (Table 1).
Besides, this moderate metallicity combined to a blue
horizontal branch,
indicates an old age for these clusters, similarly to
the RR Lyrae results by Lee (1992).  

The  bulge metallicity distribution, based
on FLAMES+GIRAFFE high resolution spectroscopy of 800 stars by Zoccali et al. (2008), 
shows that the metallicity of NGC 6522 corresponds to the lower end of the
distribution.

Abundance ratios in NGC 6522 show enhancements of the $\alpha$-elements
O, Mg and Si, whereas Ti and Ca enhancements are shallower. 
 This pattern is  shared by the similar clusters HP~1 and NGC 6558.
The shallow Ca and Ti abundances in these three clusters differs from
somewhat higher values in halo clusters, that is found as well 
in the abundances of the central metal-poor cluster Terzan 4,
as shown in Fig. \ref{pattern}. The Eu excesses are compatible among
the three clusters and as well with halo clusters. The high abundances
of the s-elements La and Ba and not found in HP~1 and NGC 6558. These high
values and the star-to-star variation are puzzling, and might point to
internal contamination from nearby AGB stars. 

In the present paper other possible members  of this class of clusters
showing moderate metallicity around [Fe/H]$\approx$-1.0 and 
blue Horizontal Branch are reported, and it would be of great interest to study them
by means of high resolution spectroscopy.

\begin{acknowledgements}
BB and EB acknowledge grants from CNPq and Fapesp.
DM acknowledges support from the FONDAP Center for Astrophysics
15010003, the BASAL Center for Astrophysics and Associated Technologies
PFB06, and FONDECYT Project 1090213.
SO acknowledges the Italian Ministero dell'Universit\`a e della Ricerca
Scientifica e Tecnologica (MURST), Italy. 
This publication makes use of data products from the Two Micron All Sky Survey,
 which is a joint project of the University of Massachusetts and the Infrared
Processing and Analysis Center/California Institute of Technology, funded by 
the National Aeronautics and Space Administration and the National Science
Foundation. 
\end{acknowledgements}

%--------------------------------- References -------------

\begin{table*}
\caption[10]{Abundance ratios derived and atomic parameters adopted.
Column 5 reports the log gf adopted  as described in Barbuy et al. (2007).
For the heavy elements BaII, LaII and EuII, hyperfine structure was taken into account,
and the log gf values reported correspond to the equivalent total log gf, adopted
from a Hill et al. (2002), b Rutten (1978), c Lawler et al. (2001b).
}
\begin{flushleft}
\begin{tabular}{llllrrrrrrrrrrrrrrrrr}
\noalign{\smallskip}
\hline
\noalign{\smallskip} 
Species & ${\rm \lambda}$     & ${\rm \chi_{ex}}$ &  C$_6$ & loggf & \multispan8 [X/Fe]    \\
 \cline{6-13}  \\
   & (${\rm \AA}$) & (eV)  & (cm$^6$s$^{-1}$) &  & {\bf B-8}  &   {\bf B-107} & {\bf B-108}   & {\bf B118} & {\bf B-122}
 &  {\bf B-128} & {\bf B-130} &  {\bf F-121} \\
\noalign{\smallskip} 
\hline
\noalign{\smallskip}
%                                                           8     107     108       118      122   128     130    121
OI & 6300.311 & 0.00 &0.30E-31 & --9.716                & +0.3: & ---     & ---    & ---    & ---   & --- & +0.5:& +0.5: \\ 
OI & 6363.79 & 0.00 &0.30E-31 & --10.25                 & +0.2: & +0.5:   & +0.7:  & +0.3: & +0.7: & ---  & ---  & +0.5: \\
NaI & 6154.230 & 2.10 & 0.90E-31 & -1.56            & +0.3  & -0.3:   &  0.0   & +0.1  & +0.1 & +0.1 & +0.1 & -0.1 \\
NaI & 6160.753 & 2.10 & 0.30E-31  &  -1.26          & +0.4  & -0.3:   & -0.2   & +0.1  & +0.2 & +0.1 & +0.2 & -0.1 \\
MgI & 6318.720 & 5.11 & 0.30E-31 & -2.10            & +0.1: & +0.20  & +0.4:  & +0.2   & +0.2 & +0.2 & +0.4 & +0.4 \\
MgI & 6319.242 & 5.11 & 0.30E-31  & -2.36           & +0.1: & ---    & +0.2:  &  0.0   & +0.2 & +0.2 & +0.4 & +0.4 \\
MgI & 6319.490 & 5.11 & 0.30E-31   & -2.80          & +0.1: & +0.20  & +0.4:  & +0.2   & +0.2 & +0.4 & +0.4 & +0.4 \\
MgI & 6765.450 & 5.75 & 0.30E-31 &  -1.94           & ---   & +0.40: & ----   & ---    & ---- & +0.2:& +0.4:& +0.4: \\
SiI & 6142.494 & 5.62 & 0.30E-31  & -1.50           & +0.2  & +0.2   & +0.4   & +0.2   &  0.0 & +0.2 & +0.4 & +0.2 \\
SiI & 6145.020 & 5.61 & 0.30E-31 & -1.45            & +0.4  & +0.2   & +0.2   & +0.2   & +0.2 & +0.2 &  --- & +0.2 \\
SiI & 6155.142 & 5.62 & 0.30E-30 & -0.85            & +0.2  & +0.1   &  0.0   & +0.2   & -0.1 & +0.2 & +0.3 & +0.2 \\
SiI & 6237.328 & 5.61 & 0.30E-30  &  -1.01          & +0.4  & +0.2   & +0.2   & +0.3   & +0.2 & +0.3 & +0.4 & +0.4 \\
SiI & 6243.823 & 5.61 & 0.30E-32  & -1.30           & +0.4  & +0.2   & +0.2   & +0.4   & +0.3 & +0.4 & +0.4 & +0.4 \\
SiI & 6414.987 & 5.87 & 0.30E-30   & -1.13          & +0.4: & +0.2   & +0.2   & +0.4   & +0.2 & +0.2 & +0.2 & +0.2 \\
SiI & 6721.844 & 5.86 & 0.90E-30 & -1.17            & +0.4  & +0.3   & +0.2   & +0.3   &  --- & +0.2: & +0.4:&  --- \\
CaI & 6156.030 & 2.52 & 4.0E-31  & -2.39 &  ---  & +0.3   & +0.1  &  0.0   &  0.0 & +0.3 & ---  &  0.0: \\
CaI & 6161.295 & 2.51 & 4.0E-31  & -1.02 & +0.2  &  0.0   & +0.3  &  0.0   & +0.3 & +0.2 & +0.3 & +0.3 \\
CaI & 6162.167 & 1.89 & 3.0E-31  & -0.09 &  0.0  & +0.2   &  0.0  & +0.2   &  0.0 &  0.0 & +0.1 & +0.3 \\
CaI & 6166.440 & 2.52 & 3.97E-31 & -0.90 &  0.0  &  0.0   & +0.2  & +0.2   & +0.3 &  0.0 & +0.3 & +0.3 \\
CaI & 6169.044 & 2.52 & 3.97E-31 & -0.54 & +0.3  & +0.1   & +0.1  & +0.2   & +0.3 & +0.3 & +0.1 & +0.3 \\
CaI & 6169.564 & 2.52 & 4.0E-31  & -0.27 & +0.1  & +0.2   & +0.1  & +0.3   & +0.3 & +0.3 & +0.2 & +0.3 \\
CaI & 6439.080 & 2.52 & 3.4E-32  & +0.3  &  0.0  & +0.3   & +0.3  & +0.3   & +0.3 & +0.3 & +0.3 &  0.0 \\
CaI & 6455.605 & 2.52 & 3.39E-32 & -1.35     & +0.3  &  0.0   & +0.3  & +0.3   & +0.3 & -0.1 & +0.4:& +0.3 \\
CaI & 6464.679 & 2.52 & 3.4E-32  & -2.10     & +0.3  &  0.0   & +0.3  & +0.3   & +0.3 & +0.3:&  --- & +0.3 \\
CaI & 6471.668 & 2.52 & 3.39E-32 & -0.59     & +0.3  &  0.0   & +0.3  & +0.3   & +0.3 & +0.3 & +0.3 & +0.3 \\
CaI & 6493.788 & 2.52 & 3.37E-32    & 0.0    & +0.1  & -0.3   &  0.0  & +0.2   &  0.0 &  0.0 & +0.1 & -0.2 \\
CaI & 6499.654 & 2.52 & 3.37E-32    & -0.85  & +0.2  & -0.3   & +0.2  &  0.0   &  0.0 &  0.0 & +0.1 & -0.1 \\
CaI & 6572.779 & 0.00 & 1.75E-32    & -4.32  & +0.1  &  0.0   & +0.3  & +0.3   & +0.2 &  0.0 & +0.2 &  0.0 \\
CaI & 6717.687 & 2.71 & 4.1E-31     & -0.61  &  0.0  &  0.0   &  0.0  & +0.3   & +0.3 & +0.3 & +0.3 & +0.1\\
TiI & 6126.224 & 1.07 & 1.37E-32    & -1.43  & +0.2  & +0.2   & +0.2  &  0.0   & +0.1 & +0.1 & +0.2 &  0.0 \\
TiI & 6258.110 & 1.44 & 3.17E-32    & -0.36  &  0.0  &  0.0   & +0.2  &  0.0   & +0.2 & +0.2 &  0.0 & +0.1 \\
TiI & 6261.106 & 1.43 & 4.68E-32     & -0.48  &  0.0  &  0.0   &  0.0  &  0.0   & +0.3 &  0.0 &  0.0 & +0.2 \\
TiI & 6303.767 & 1.44 & 3.12E-32    & -1.57  &  0.0: &  0.0   &  ---  & +0.2   & +0.2:& +0.2:  &  --- & ---  \\
TiI & 6336.113 & 1.44 & 1.86E-32    & -1.74  &  ---  &  ---   & +0.2: & +0.3   & +0.3 &  0.0: &  --- & +0.3: \\
TiI & 6554.238 & 1.44 & 1.81E-32    & -1.22  & +0.2  & +0.2   & +0.2  &  0.0   & +0.2 & +0.2 &  0.0 & +0.2 \\
TiI & 6556.077 & 1.46 & 1.81E-32    & -1.07  &  0.0  & +0.2   & +0.3  &  0.0   &  0.0 & +0.2 & +0.4 & +0.2 \\
TiI & 6599.113 & 0.90 & 1.96E-32    & -2.09  & +0.2: & +0.3   & +0.2: & +0.3   & +0.2 & +0.2::& +0.4: & +0.1 \\
TiI & 6743.127 & 0.90 & 0.30E-31               & -1.73  &  0.0  & +0.1   & +0.2  & -0.2   &  0.0 &  0.0 & +0.2 & +0.2 \\
TiII& 6491.580 & 2.06 & 0.30E-31               & -2.10  & +0.2  & +0.3   & +0.3  & +0.3   & +0.3 & +0.3 & +0.3 & +0.1 \\
TiII& 6559.576 & 2.05 & 0.30E-31               & -2.35  & +0.3  &  0.0   & +0.3  & +0.3   & +0.3 & +0.3 & +0.3 & +0.2 \\
TiII& 6606.970 & 2.06 & 0.30E-31               & -2.85  & +0.2  & +0.2:  & +0.2  & ---    & ---  & +0.3:& +0.3 & +0.2 \\
BaII & 6141.727 & 0.70 & & -0.07$^a$                          & +1.0 & +0.5   & -0.1  & +0.75   & +0.5 & +1.0 & +0.25 &  -0.25  \cr
BaII & 6496.908 & 0.60 & & -0.38$^b$                          & +0.9 & +0.5   & +0.15  & +1.2   & +0.75 & +0.75 & +0.25 &  -0.25 \cr
LaII & 6390.480 & 0.32 & & -1.41$^c$                      & +0.5 & +0.50   & +0.3  & +0.5   & +0.3:&  --- & ---  &  0.0:\cr
EuII & 6645.127 & 1.38 & & 0.12$^a$                      & +0.5 & 0.0:  &  +0.5:  & +0.5   & +0.3:&  0.0: & +0.8 & +0.5 \cr
\noalign{\smallskip} \hline \end{tabular}
\label{tablines}
\end{flushleft}
\end{table*}

%===============================================================================
\begin{table*}
\begin{flushleft}
\caption{Final abundances for each sample star, and mean results and corresponding internal errors.}             
\label{final}      
\centering          
\begin{tabular}{cccccccccccc}     % 12 columns 
\noalign{\smallskip}
\hline\hline    
\noalign{\smallskip}
\noalign{\vskip 0.1cm} 
Element & B-8 & B-107 & B-108 & B-118 & B-122 & B-128 & B-130 & F-121 & Mean & \\                 
\noalign{\vskip 0.1cm}
\noalign{\hrule\vskip 0.1cm}
\noalign{\vskip 0.1cm}    
%\multicolumn{7}{c}{\hbox{Spectroscopic}}  \\ 
\noalign{\vskip 0.1cm}
\noalign{\hrule\vskip 0.1cm}    
\hbox{[O/Fe]}  & +0.25:  & +0.5:  & +0.7:  & +0.3: & +0.7: &  ---  & +0.5: & +0.5: & +0.39:$\pm$0.3 & \\
\hbox{[Na/Fe]} & +0.35  & -0.30  & -0.15   &  +0.10  & +0.15  & +0.10  & +0.15  & -0.10  & +0.03$\pm$0.3 &\\
\hbox{[Mg/Fe]}  & +0.10  & +0.27  & +0.33   & +0.20 & +0.20 & +0.25  & +0.40  & +0.40  & +0.27$\pm$0.15 &\\
\hbox{[Si/Fe]}  & +0.34 & +0.20  &  +0.20   & +0.29 & +0.13 & +0.24 & +0.35 & +0.27  & +0.25$\pm$0.10 &\\
\hbox{[Ca/Fe]}  & +0.15 & +0.04 & +0.18  & +0.21 & +0.21 & +0.16 & +0.23 & +0.16 & +0.17$\pm$0.06 &\\
\hbox{[Ti/Fe]}  & +0.12 & +0.14 & +0.21  & +0.11  & +0.19 & +0.17 & +0.21 & +0.16 & +0.16$\pm$0.04 &\\
\hbox{[Eu/Fe]}  & +0.50 & \phantom{-}0.00 & +0.50  & +0.50   & +0.30 & \phantom{-}0.00 & +0.80 & +0.50 & +0.39$\pm$0.4 &\\
\hbox{[Ba/Fe]}  & +0.95 & +0.50 &  $\phantom{-}$0.0  & +1.00 & +0.60 & +0.90 & +0.25 & -0.25   & +0.49$\pm$0.4 &\\
\hbox{[La/Fe]}  & +0.50 & +0.50 & +0.30  & +0.50 &  +0.30  & ---   & ---   &  0.00   & +0.35$\pm$0.2 &\\
\hline                  
\end{tabular}
\end{flushleft}
\end{table*}
%===============================================================================

\begin{table}
\caption[9]{Final abundance ratios [X/Fe] of NGC 6522,
compared with those of the metal-poor bulge clusters
HP~1 (Barbuy et al. 2006), NGC 6558 (Barbuy et al. 2007),
 UKS 1 (Origlia et al. 2005) and Terzan 4 (Origlia et al. 2004).
 In the 2nd column the solar abundances adopted are reported.}
%\begin{flushleft}
\tabcolsep 0.13cm
\begin{tabular}{lcrrrrrrrrrrr}
\noalign{\smallskip}
\hline
\noalign{\smallskip}
Species & $\epsilon$(X)$_{\odot}$ &  NGC 6522 &  NGC 6558 & HP~1 & UKS 1  & Terzan 4 \\
\hline
\noalign{\smallskip}
\hline
\noalign{\smallskip}
[Fe/H]   & 7.50     & --1.00 & --0.97 & --0.99 & -0.78 &  -1.60 & \cr
[OI/Fe]  & 8.77     &  +0.39 &  +0.38 &  +0.40 & +0.27 &  +0.55 & \cr
[NaI/Fe] & 6.33     &  +0.03 & --0.09 &  +0.00 & --- & ---  & \cr
[MgI/Fe] & 7.58     &  +0.27 &  +0.24 &  +0.10 & +0.32 &   +0.41 &  \cr
[SiI/Fe] & 7.55    &  +0.25 &  +0.23 &  +0.30 & +0.28 &  +0.55 & \cr
[CaI/Fe] & 6.36   &  +0.17 &  +0.05 & +0.03 & +0.38 &  +0.53 & \cr
[TiI,II/Fe] & 5.02   & +0.16  &  +0.06 &  +0.08 &  +0.32 &  +0.43 & \cr
[BaII/Fe]& 2.13   & +0.49  & +0.13 &  +0.15 &   --- & --- \cr
[LaII/Fe]&1.22   &  +0.35   & 0.00   & +0.00 &    --- & --- \cr
[EuII/Fe]&0.51    & +0.39  & +0.36 &  +0.15 &  --- & --- \cr
\noalign{\smallskip} \hline \end{tabular}
\label{tabalpha}
\end{table}

%===============================================================================
%                          Online Material
%===============================================================================
\Online
\appendix
\begin{longtable}{l@{}crrrrr@{}r@{}r@{}r@{}r@{}r@{}r@{}r@{}r@{}r}
\caption{\label{gf}Fe I and FeII line list, 
wavelength, excitation potential, damping constant,  gf-values and equivalent widths.
The oscillator strengths used in this paper and in Barbuy et al. (2007) are given in column
6, and compared to those given in Fuhr \& Wiese (2006) in column 5.
Equivalent widths were measured using DAOSPEC. For B-108 and B-134 the lines
were checked with IRAF, and only very clean lines were kept. }\\
\noalign{\smallskip}
\hline
\noalign{\smallskip}
\hbox{Species} & \hbox{${\rm \lambda}$} & \hbox{${\rm \chi_{ex}}$} &
\hbox{${\rm C_6}$} & \hbox{log $gf$} & \hbox{log $gf$} & \hbox{B-8} & 
\hbox{\phantom{-}B-107} &
 \hbox{\phantom{-}B-108} & \hbox{\phantom{-}B-118} & \hbox{\phantom{-}B-122} &
 \hbox{\phantom{-}B-128} & \hbox{\phantom{-}B-130} & \hbox{\phantom{-}B-134} 
 & \hbox{\phantom{-}F-121} & \\
 & \hbox{({\rm \AA})} & \hbox{(eV)} &
\hbox{${\rm cm^6s^{-1}}$} & \hbox{FW06} & \hbox{present} &  &  & &  &  &  &  &  &  & \\
\noalign{\hrule\vskip 0.1cm}
\noalign{\vskip 0.1cm}
%\hbox{Fe II}  & 6084.12 & 3.20 & 3.0E$-$32 & -3.81  &        &  &  & & & & & & & & & \\
FeII &  6149.26 & 3.89  & 0.34E-32  & -2.84  & -2.69 &  18 &  31 &  22 &  23 &  25 &  31 &  32 &  40 &  20 &  \\
FeII &  6247.56 & 3.89  & 0.33E-32  & -2.43  & -1.98 &  43 &  51 &  25 &  43 &  42 &  44 &  43 &  57 &  43 & \\
FeII &  6432.68 & 2.89  & 0.24E-32  & -3.57  & -3.57 &  35 &  42 &  28 &  29 &  33 &  38 &  38 &  38 &  30 & \\
FeII &  6456.39 & 3.90  & 0.32E-32  & -2.19  & -2.05 &  60 &  76 &  49 &  62 &  53 &  51 &  53 &  85 &  47 & \\
FeII &  6516.08 & 2.89  & 0.25E-32  & -3.37  & -3.31 &  52 &  53 &  34 &  40 &  44 &  44 &  43 &  60 &  49 & \\
FeI &   6151.62 & 2.18  & 0.81E-32  & same   & -3.299&  73 &  44 &  52 &  74 &  59 &  68 &  52 &  27 &  56 & \\
FeI &   6159.38 & 4.61  & 0.13E-30  & absent & -1.97 &  24 & --- & --- & --- & --- & --- &  13 & --- & --- & \\
FeI &   6165.36 & 4.14  & 0.77E-31  & -1.474 & -1.549&  32 &  23 & --- &  37 & --- &  29 &  25 &   7 &  25 & \\
FeI &   6173.34 & 2.22  & 0.84E-32  & -2.880 & -2.879&  95 &  67 &  62 &  99 &  88 &  92 &  78 & --- &  75 &\\
FeI &   6180.20 & 2.73  & 0.13E-31  & -2.650 & -2.784&  78 &  44 &  50 &  68 &  58 &  70 &  60 &  12 &  54 & \\
FeI &   6187.99 & 3.94  & 0.30E-30  & -1.670 & -1.718&  40 &  22 &  21 &  42 &  27 &  34 &  31 &  19 &  26 & \\
FeI &   6200.31 & 2.61  & 0.15E-31  & same   & -2.437&  83 &  64 &  66 &  81 &  73 &  81 &  76 &  11 &  75 & \\
FeI &   6213.43 & 2.22  & 0.30E-31  & -2.48  & -2.646& 109 &  87 &  78 &  99 &  96 &  98 &  91 &  35 &  87 & \\
FeI &   6220.78 & 3.88  & 0.13E-30  & absent & -2.46 &  22 &  11 & --- &  16 & --- & --- & --- & --- &  12 &\\
FeI &   6226.74 & 3.88  & 0.13E-30  & absent & -2.202&  23 &  16 & --- &  15 &  16 &  17 &  15 &  17 &  17 & \\
FeI &   6240.65 & 2.22  & 0.10E-31  & -3.17  & -3.388&  71 &  86 &  45 &  68 &  57 &  73 &  50 & --- &  56 & \\
FeI &   6246.32 & 3.60  & 0.12E-30  & -0.877 & -0.956& 102 &  86 &  73 & 102 &  95 & 110 & 100 &  64 &  99 & \\
FeI &   6252.56 & 2.40  & 0.12E-31  & same   & -1.687& 140 & 117 & 103 & 126 & 116 & 118 & 119 &  82 & 119 & \\
FeI &   6254.25 & 2.28  & 0.13E-31  & -2.426 & -2.480& 115 & 103 &  87 & 127 & 108 & 116 & 106 &  52 & 106 & \\
FeI &   6270.23 & 2.86  & 0.15E-31  & -2.61  & -2.711&  73 &  36 &  51 &  67 &  60 &  72 &  62 &  21 &  56 & \\
FeI &   6271.28 & 3.32  & 0.89E-31  & -2.703 & -2.957&  17 &   9 &  18 &  21 & --- &  12 & --- &  13 &  14 & \\
FeI &   6297.79 & 2.22  & 0.82E-32  & same   & -2.740&  97 &  58 &  63 &  88 &  75 &  90 &  72 &  25 &  71 &\\
FeI &   6301.50 & 3.65  & 0.23E-31  & -0.718 & -0.720& 100 &  90 &  67 &  89 &  84 &  86 &  90 & --- &  86 &\\
FeI &   6302.50 & 3.69  & 0.23E-31  & absent & -0.91 &  91 &  75 &  66 &  70 &  72 &  88 &  79 &  37 &  73 &\\
FeI &   6311.50 & 2.83  & 0.14E-31  & -3.141 & -3.224&  46 &  28 &  38 &  45 &  43 &  46 &  44 &  32 &  25 & \\
FeI &   6315.31 & 4.14  & 0.30E-31  & -1.232 & -1.230&  55 &  42 &  33 &  52 &  42 &  48 &  33 & --- &  43 & \\
FeI &   6315.81 & 4.08  & 0.66E-31  & -1.660 & -1.712&  38 &  30 &  30 &  41 &  42 &  39 &  30 & --- &  23 & \\
FeI &   6322.69 & 2.59  & 0.14E-31  & same   & -2.426&  92 &  69 &  59 &  91 &  77 &  89 &  70 &  28 &  78 & \\
FeI &   6335.33 & 2.20  & 0.80E-32  & -2.177 & -2.229& 114 & 106 &  90 & 109 & 110 & 120 & 103 & --- &  99 & \\
FeI &   6336.82 & 3.69  & 0.13E-30  & -0.856 & -1.053& 101 &  77 &  70 &  90 &  80 & --- &  69 &  36 &  84 & \\
FeI &   6344.15 & 2.43  & 0.12E-31  & -2.923 & -2.922&  93 &  56 &  55 &  78 &  75 &  84 &  65 &  29 &  70 & \\
FeI &   6355.03 & 2.84  & 0.30E-31  & -2.291 & -2.40 &  97 &  35 &  58 &  89 &  80 &  94 &  70 &  24 &  74 & \\
FeI &   6392.54 & 2.28  & 0.11E-31  & absent & -4.03 &  42 & --- & --- &  31 &  18 &  24 &  31 & --- &  21 & \\
FeI &   6393.60 & 2.43  & 0.12E-31  & -1.58  & -1.615& 183 & 136 & 107 & 144 & 123 & 152 & 143 & --- & 107 & \\
FeI &   6419.95 & 4.73  & 0.15E-30  &-0.27   & -0.250&  64 &  41 &  47 &  66 &  53 &  62 &  53 &  28 &  52 & \\
FeI &   6430.85 & 2.18  & 0.77E-32  & -2.006 & -2.005& 133 & 112 & 116 & 133 & 121 & 133 & 124 &  74 &  97 & \\
FeI &   6475.62 & 2.56  & 0.15E-31  & same   & -2.94 &  77 &  42 &  57 &  77 &  57 &  64 &  61 &  17 &  44 & \\
FeI &   6481.87 & 2.28  & 0.11E-31  & same   & -2.984&  87 &  58 &  64 &  81 &  72 &  78 &  74 &  25 &  55 & \\
FeI &   6498.94 & 0.96  & 0.49E-32  & -4.687 & -4.699&  87 &  44 &  74 &  71 &  68 &  73 &  70 &  15 &  51 & \\
FeI &   6518.37 & 2.83  & 0.13E-31  & -2.30  & -2.748&  61 &  46 &  45 &  50 &  60 &  61 &  48 &  19 &  50 & \\
FeI &   6533.93 & 4.56  & 0.16E-30  & -1.43  & -1.453&  20 &  19 & --- & --- &  25 & --- &  25 & --- &  19 & \\
FeI &   6569.22 & 4.73  & 0.13E-30  & -0.45  & -0.422&  59 &  38 &  49 &  62 &  52 &  69 &  62 & --- &  45 & \\
FeI &   6574.23 & 0.99  & 0.56E-32  & -5.004 & -5.042&  63 &  18 &  41 &  64 &  49 &  50 &  45 & --- &  51 & \\
FeI &   6575.02 & 2.59  & 0.14E-31  & -2.71  & -2.824&  85 &  47 &  57 &  78 &  69 &  79 &  64 & --- &  62 & \\
FeI &   6591.31 & 4.59  & 0.12E-30  & absent & -2.06 & --- & --- & --- & --- & --- &   9 & --- & --- & --- & \\
FeI &   6593.87 & 2.43  & 0.12E-31  & same   & -2.422& 105 &  86 &  86 & 115 &  98 & 105 &  87 & --- &  80 & \\
FeI &   6597.56 & 4.80  & 0.15E-30  & -1.05  & -1.061&  33 &  13 &  21 &  28 &  31 &  34 &  17 &  19 &  25 & \\
FeI &   6608.03 & 2.28  & 0.10E-31  & absent & -4.038&  16 &   7 & --- &  14 &  25 &  15 &  13 &   6 &  16 & \\
FeI &   6678.00 & 2.69  & 0.30E-31  & absent & -1.420& 147 & 118 & --- & 150 & 139 & 148 & 134 & --- & 128 & \\
FeI &   6703.57 & 2.76  & 0.12E-31  & -3.06  & -3.15 &  51 &  35 &  28 &  40 &  37 &  54 &  19 &  19 &  33 & \\
FeI &   6705.11 & 4.61  & 0.30E-31  & absent & -1.060&  34 & --- & --- &  29 &  12 &  40 &  26 & --- &  22 & \\
FeI &   6710.32 & 1.48  & 0.64E-32  & absent & -4.874&  46 &  14 & --- &  49 &  31 &  46 &  32 & --- &  34 & \\
FeI &   6713.75 & 4.80  & 0.16E-30  & absent & -1.602& --- & --- & --- &  16 & --- & --- &   5 & --- &   7 &\\
FeI &   6715.38 & 4.59  & 0.11E-30  & absent & -1.638&  32 &  17 & --- &  14 &  12 &  27 &  18 & --- &  18 & \\
FeI &   6716.24 & 4.56  & 0.79E-32  & absent & -1.927&  15 &  11 & --- & --- &   8 &  31 &  10 & --- &  11 & \\
FeI &   6725.36 & 4.10  & 0.15E-30  & absent & -2.300& --- & --- & --- &   9 &  10 &   5 &  20 & --- &  16 & \\
FeI &   6726.67 & 4.59  & 0.30E-31  & absent & -1.090&  22 & --- &  27 &  24 &  34 &  18 &  32 & --- &  24 &\\
FeI &   6733.15 & 4.64  & 0.11E-30  & absent & -1.576&  25 &  22 & --- &  30 &  27 &  31 &  18 & --- &  23 & \\
FeI &   6739.52 & 1.56  & 0.67E-32  & -4.79  & -4.941&  33 & --- & --- &  26 &  36 &  25 &  25 & --- &  30 & \\
FeI &   6752.71 & 4.64  & 0.11E-30  & -1.204 & -1.366&  22 &  27 & --- &  23 &  28 &  23 &   6 & --- &  21 & \\
FeI &   6810.26 & 4.60  & 0.14E-30  & -0.986 & -1.108&  31 &  11 & --- &  44 &  25 &  35 &  32 & --- &  17 & \\
FeI &   6820.37 & 4.64  & 0.16E-30  & -1.29  & -1.311&  36 &  12 & --- &  45 &  23 &  30 &  14 & --- &  24 & \\
FeI &   6837.02 & 4.59  & 0.78E-32  & -1.687 & -1.800&  13 & --- & --- &  22 & --- & --- &  24 & --- &  14 & \\
FeI &   6839.83 & 2.56  & 0.13E-31  & -3.35  & -3.451&  49 &  24 & --- &  37 &  32 &  47 &  35 & --- &  27 & \\
FeI &   6841.34 & 4.61  & 0.10E-30  & -0.78  & -0.752&  49 &  27 & --- &  56 &  53 &  55 &  17 & --- &  39 & \\
FeI &   6842.69 & 4.64  & 0.15E-30  & -1.29  & -1.315&  20 &  11 & --- &  23 &  44 &  38 & --- & --- &  13 & \\
FeI &   6843.66 & 4.55  & 0.94E-31  & -0.89  & -0.928&  50 &  23 & --- &  47 &  53 &  64 &  32 & --- &  42 & \\
FeI &   6851.63 & 1.60  & 0.73E-32  & absent & -5.307&  26 &  21 & --- &  18 &  20 &  30 &   8 & --- & --- & \\
FeI &   6855.71 & 4.39  & 0.10E-30  & absent & -1.819&  33 &  35 & --- &  45 &  17 &  28 &  25 & --- &  28 & \\
FeI &   6857.25 & 4.08  & 0.53E-31  & absent & -2.156&  15 &  17 & --- &  14 &  13 &  24 &   8 & --- &  21 & \\
FeI &   6858.15 & 4.61  & 0.10E-30  & -0.93  & -1.055&  39 &  30 & --- &  30 &  35 &  41 &  13 & --- &  24 & \\
%\noalign{\smallskip} \hline \end{tabular}  
\end{longtable}

\end{document}